\documentclass[]{aa}

\usepackage{graphicx}

\newcommand{\Msol}{\mbox{M$_{\odot}$}}              

\begin{document}

\title{Cool carbon stars in the halo : new very red or distant objects
\thanks{Based on observations done at  Haute Provence Observatory
 operated by the Centre National de Recherche Scientifique (France),
 and at Byurakan Observatory (Armenia)
}}

\author{N.~Mauron\inst{1} \and  K.S.~Gigoyan\inst{2} \and T.R.~Kendall\inst{3}} 

\offprints{N.Mauron}

\institute{GRAAL, CNRS and Universit\'e  Montpellier II,  
 Place Bataillon, 34095 Montpellier, France\\ 
 \email{mauron@graal.univ-montp2.fr}
\and 
378433 Byurakan Astrophysical Observatory and Isaac 
        Newton Institute of Chile, Armenian Branch, Ashtarak d-ct, Armenia\\
\email{kgigoyan@bao.sci.am}
\and 
Centre for Astrophysics Research, Science and Technology Research Institute,
University of Hertfordshire, College Lane, Hatfield AL10 9AB, United Kingdom\\
\email{trkendall@googlemail.com} }

\date{Received xx ; accepted xx }

\abstract{}
{The goal of this  paper is to present and analyse a new sample
of cool carbon (C) stars located in the Galactic halo.} 
{These rare objects are discovered  by searching the 2MASS  point-source catalogue for 
candidates having near-infrared colours typical of C stars. Optical 
spectroscopy is subsequently performed.}
{Twenty-three new C stars were discovered, with their $K_{\rm s}$ magnitude 
in the range 6 to 13.3. 
Spectra are typical of N-type carbon stars with C$_{2}$ and CN bands 
and sometimes H$\alpha$ in emission. One object is a S-type star. When the objects 
are bright enough ($V \le 15.5$), the data of the Northern Sky Variability Survey can be 
exploited. In all cases, stars belonging to this survey  show light variations 
confirming that they are AGB stars. Distances and galactocentric $XYZ$ coordinates  
have been estimated by assuming these stars to be similar in luminosity to those of the 
Sagittarius dwarf galaxy. Four objects are particularly red with  $J$-$K_{\rm s} > 3$, 
with two located at more than 5 kpc from the Galactic  plane. Eight additional objects  
with similar properties are found in the litterature and our previous works. 
These 12 C stars could be useful to study mass loss at low metallicity.
 Two other objects are remarkably far from the Sun, at distances of 95 and 110\,kpc. They
are located, together with two other C stars previously found,
in the region $Z < -60$\,kpc in which the model of Law et al. (2005) predicts
 the Sgr Stream to have a loop.} 
{}

\titlerunning{Cool carbon stars in the halo}
\authorrunning{N.~Mauron, K.S.~Gigoyan \& T.R.~Kendall}
\maketitle


\section{Introduction}

   Cool carbon (C) stars are objects  evolving on the asymptotic giant 
   branch (AGB). This is a  brief phase of stellar evolution characterized 
   by high luminosity, low effective temperature, and  the appearance at the 
   surface of the star of nuclear products fabricated in its internal layers
   (for  reviews on AGB and C stars, see Habing \cite{habing96}, Wallerstein 
   \& Knapp \cite{walknapp98})
   The  progenitors of these C stars have  intermediate masses, from about 
   1.3\,\Msol\, in most cases,  up to about 4\,\Msol\,.
   C stars are   variable sources. They   can be recognized  by their 
   very red colours and  the strong spectral features arising
   from   carbon-rich molecules. Because evolution on the AGB is fast,  
   they  are not very abundant in the Galactic disc. However, because of 
   their  luminosity, they  have been  detected and studied in many galaxies 
   of the Local Group and beyond. They are more numerous in systems
  with an average low metallicity, and generally indicate the presence of an 
  intermediate age population (see for example Wallerstein \& Knapp 
  \cite{walknapp98}; Van den Bergh \cite{vandenbergh00}; Mouhcine \& Lan\c{c}on 
  \cite{mouhcine03}; Groenewegen \cite{groe05}). 
  
  In the Galactic disc and close to the Sun, i.e. at distances less than 
  $\sim$\,5\,kpc, AGB C stars have a scale height  of 200\,pc
  (Claussen et al. \cite{claussen87}; Groenewegen et al. \cite{groe92}; 
   Bergeat et al. \cite{bergeat02}).
   Therefore,  very few  of them are expected at  more than about 
   $\sim$\,1\,kpc 
  above or below the Galactic  plane. In the old halo population, all stars with masses 
  similar to those of C stars have evolved beyond the luminous AGB phase. 
  However, it has long been known that  some  faint and 
  cool carbon stars do exist at large distances and at high galactic latitude.
   Totten and Irwin (\cite{ti98}, hereafter TI98) listed 41  previously known objects,
  this number including some warmer CH-type objects. 
  They  discovered 36 additional cases, with 29  N-type and 7 CH-type (for more
  information about classification of C stars, see  Wallerstein \& Knapp \cite{walknapp98}). 
  One of  their goals was to show that these halo C stars might arise from 
  the disruption  of dwarf galaxies captured by the Galaxy. This was 
  eventually proved by Ibata et al. (\cite{ibata01}) who found that more than half of 
  these C stars are coherently clustered on a great circle on the sky, and 
  trace the debris stream of the Sagittarius dwarf galaxy.  
  
  The goal of this paper is to present new discoveries of halo cool C stars
  and to study their properties.
  In our previous works  (Mauron et al. \cite{mauron04}, \cite{mauron05}), we  
  followed the approach of TI98 to search for  these rare
  halo cool C stars.  However, rather than using visible photometry from 
  photographic plates to select candidates, 
  we have exploited   the recent 2MASS near-infrared survey
  and its point-source catalogue (Cutri et al. \cite{cutri03}).
  This survey  provides us with accurate $JHK_{\rm s}$ photometry, 
  enabling the potential discovery of C stars {\it via} their  near-infrared colours.
  Follow-up spectroscopy is then  performed to determine whether these candidates are
  carbon-rich, and show the expected signatures of evolved AGB objects. 
  
  A first result of our research was that 
  50 new cool halo  C stars were discovered.
  This suggested that, firstly, the 
  observational strategy was correct (for details on the candidate selection process,
  see Mauron et al. \cite{mauron05}). Secondly, it was found that a majority
  of  these objects belong to the Sagittarius (Sgr) stream, but not all of them do,
  in agreement with previous findings. Finally, we found several  remarkably 
  faint ($K_{\rm s} \ge 11$) objects;  interesting probes
  of the halo at distances from the Sun as large as  $\sim$ 60-120 kpc.  
  
  In this paper, we report on our continuing program to enlarge the sample 
  of these halo C stars. We wish  eventually to establish a complete 
  $K_{\rm s}$-band limited inventory. Two new very distant sources 
  were discovered with $K_{\rm s} > 12$. 
  Observations  and data reduction are described in Sect.\,2.
  In Sect.~3 we analyse our new sample  of 25 halo C stars, and we discuss 
  global properties of the sample and individual
  cases, including consideration of colours, spectra and variability. Very red
  sources are discussed in Sect.~4.  In Sect.~5, 
  we estimate the distances of the sources and  discuss their location 
  in the Galactic  halo. A summary of this work and our conclusions are finally
  given  in Sect.\,6. 
  
 
 \section{Observations}
Slit spectroscopy of our candidates was performed at the Byurakan Observatory
(Armenia) and at Haute-Provence Observatory (France). At Byurakan, the 
observations were carried out at various dates during 2005. 
The instrument was the SCORPIO spectrograph attached to the 2.6-m 
telescope. The spectrograph was used with a 600 g mm$^{-1}$  grating and a
Lick-3 2064 x 2058 CCD detector which has 15$\times$15 $\mu$m 
pixels. The resulting dispersion is 1.7\,\AA\, per pixel. The resolving power 
is $\lambda/\delta\lambda$\,=\,750, hence $\delta\lambda$\,=\,8\,\AA\, at 6000\AA\,.
The useful spectral region is from 4600\,\AA\, to 7000\,\AA.

 At Haute-Provence, the observations were made during the nights
 September 5 to 9, 2005 and September 27 to October 2, 2006.
 The instrument was the CARELEC spectrograph mounted 
 at the Cassegrain focus of the 1.93-m telescope. The spectrograph 
 was used with a 150 g mm$^{-1}$ grating and an EEV 2048$\times$1024 
 CCD chip with 13.5$\times$13.5 $\mu$m pixels. 
 The dispersion  is 3.6\,\AA\, per pixel. The slit width was 
 2.0 arcsec. The resolving power is $\lambda/\delta\lambda$\,=\,460, and 
 $\delta\lambda$\,=\,13\,\AA\,at 6000\,\AA\,, with a spectral coverage of
  4400 to 8400\,\AA.

 Typical exposure times were from about 1 minute for the brightest candidate
 carbon stars ($R \sim 12$) to about 1 hour for the faintest ones 
 ($R \sim 17$). The low spectral resolution was chosen to allow
 spectroscopy of these faint stars in a reasonable time, and the
 spectra  have sufficient signal to noise ratio ( $\geq$ 20) to 
 perfectly recognize 
 the strong molecular features of carbon-rich objects. Most of the
 contaminants in our list of C candidates are  cool M giants or M dwarfs. 
 Faint sources required clear conditions
 with good seeing. When these conditions were not fulfilled,  only bright 
 sources were observed. 
 
 Standard data reduction for slit spectroscopy was carried out with 
 the ESO Midas software. It included bias subtraction, flat-fielding,
 extraction of one-dimensional spectra for the object and for the sky, 
 subtraction of the sky spectrum, cleaning of cosmic rays, and
 wavelength calibration. Then,
 a correction was applied to take into account the instrumental spectral 
 efficiency, and this was done with the spectrum of a standard 
 photometric star observed  with the same instrumental set-up. 
 The spectra are thus proportional to a flux expressed in
 erg s$^{-1}$ cm$^{-2}$ \AA$^{-1}$ and can be compared to those shown 
 by TI98. However, no absolute calibration could be
 achieved because the sky was most of the time not photometric, and 
 because  strong slit losses occured when the seeing was bad.
 Therefore, only  relative spectral distributions 
 can finally be considered and are drawn in our plots.

\begin{table*}[!ht]
	\caption[]{List of observed  carbon stars and their main characteristics. The
	 columns give the running number, the 2MASS name, the galactic coordinates, 
	the $B$ and $R$ magnitudes and the $B-R$ colour index, the 
	$J H K_{\rm s}$ magnitudes from 2MASS and the $J-K_{\rm s}$ colour index.}
	\begin{flushleft}
	\begin{tabular}{lcrrrrrrrrrl}
	\noalign{\smallskip}
	\hline
	\hline
	\noalign{\smallskip}
No.&  2MASS name  & $l$~~~~ & $b$~~ & $B$ & $R$~~ & $B$-$R$ & $J$~~ & $H$~~ & $K_{\rm s}$~~ & $J$-$K_{\rm s}$&Note\\
	\noalign{\smallskip}
	\hline
 59& \object{2MASS J003504.77$+$010845.8} & 114.351 & $-$61.453 & 19.2 &16.7 & 2.5 & 14.620 &  13.734 &  13.280 &  1.340& \\
 60& \object{2MASS J011256.39$+$395945.3} & 127.393 & $-$22.689 & 15.1 &11.3 & 3.8 &  8.181 &   7.160 &   6.725 &  1.456& \\
 61& \object{2MASS J014736.29$+$371229.1} & 135.175 & $-$24.325 & 17.4 &13.3 & 4.1 & 10.523 &   9.436 &   8.884 &  1.639& \\
 62& \object{2MASS J020056.14$+$094535.6} & 149.845 & $-$49.446 & 19.7 &15.2 & 4.5 & 10.266 &   8.525 &   7.170 &  3.096& \\
 63& \object{2MASS J021012.06$-$015738.9} & 163.123 & $-$58.546 & 16.2 &12.2 & 4.0 & 11.505 &  10.605 &  10.010 &  1.495& 1\\
 64& \object{2MASS J030011.33$+$164940.2} & 162.049 & $-$36.082 & 17.5 &14.2 & 3.3 & 11.632 &  10.742 &  10.293 &  1.339&\\
 65& \object{2MASS J034828.12$+$165703.2} & 172.347 & $-$28.475 & 18.1 &14.8 & 3.3 & 13.098 &  11.903 &  11.097 &  2.001& \\
 66& \object{2MASS J035010.66$+$260502.8} & 165.713 & $-$21.582 & 16.7 &13.3 & 4.4 & 10.240 &   9.244 &   8.803 &  1.437& \\
 67& \object{2MASS J040143.35$+$084210.6} & 181.911 & $-$31.637 & 16.1 &12.6 & 3.5 &  9.765 &   8.703 &   8.195 &  1.570& \\
 68& \object{2MASS J040648.84$+$162818.3} & 176.133 & $-$25.619 & 16.9 &13.0 & 3.9 &  9.573 &   8.517 &   8.036 &  1.537& \\
 69& \object{2MASS J042638.77$+$142516.1} & 181.245 & $-$23.296 & 19.2 &17.1 & 2.1 & 13.215 &  12.030 &  11.204 &  2.011& 2\\
 70& \object{2MASS J045602.71$+$092219.1} & 190.204 & $-$20.452 & 17.3 &13.4 & 3.9 & 10.189 &   9.141 &   8.599 &  1.590& \\
 71& \object{2MASS J060634.52$+$731026.9} & 140.861 & $+$22.738 & 15.9 &12.7 & 3.2 &  9.403 &   8.393 &   7.891 &  1.512& \\
 72& \object{2MASS J064618.05$+$543133.6} & 161.381 & $+$21.086 & 14.8 &12.8 & 2.0 &  9.478 &   8.458 &   8.072 &  1.406& \\
 73& \object{2MASS J071057.47$+$475818.0} & 169.337 & $+$22.917 & 16.1 &12.3 & 3.8 &  8.935 &   7.880 &   7.352 &  1.583& \\
 74& \object{2MASS J171755.52$+$043606.9} &  26.142 & $+$22.867 & 17.3 &14.5 & 2.8 & 11.557 &  10.592 &  10.200 &  1.357& \\
 75& \object{2MASS J175815.64$+$223551.1} &  48.195 & $+$21.356 & 15.7 &12.0 & 3.7 &  9.516 &   8.632 &   8.073 &  1.443& \\
 76& \object{2MASS J193930.23$+$754140.5} & 107.672 & $+$23.332 & 16.2 &12.5 & 3.7 &  9.426 &   8.326 &   7.783 &  1.643& \\
 77& \object{2MASS J195840.16$+$774526.2} & 110.190 & $+$23.019 & 21.0 &14.9 & 6.1 & 12.048 &  10.161 &   8.692 &  3.356& \\
 78& \object{2MASS J201559.67$+$763508.4} & 109.469 & $+$21.659 & 16.9 &12.9 & 4.0 &  9.553 &   8.573 &   7.978 &  1.575& \\ 
 79& \object{2MASS J210522.23$+$780116.2} & 112.308 & $+$20.130 & 17.0 &13.0 & 4.0 &  9.779 &   7.699 &   6.116 &  3.663& \\
 80& \object{2MASS J211944.93$+$180029.7} &  68.200 & $-$21.673 & 18.1 &13.6 & 4.5 &  8.671 &   7.338 &   6.364 &  2.307& 3\\
 81& \object{2MASS J212318.24$-$140819.5} &  37.019 & $-$39.965 & 18.7 &16.3 & 2.4 & 14.281 &  13.427 &  12.862 &  1.419& \\
 82& \object{2MASS J215526.97$+$234214.4} &  78.911 & $-$23.760 & 18.3 &14.0 & 4.3 &  8.901 &   6.986 &   5.503 &  3.398& \\
 83& \object{2MASS J222301.20$+$221656.5} &  83.298 & $-$28.940 & 20.2 &15.0 & 5.2 & 11.685 &  10.392 &   9.524 &  2.161& \\

         \noalign{\smallskip}
   \hline
\end{tabular}
\end{flushleft}

{\small Notes: (1) carbon star identified by TI98. It was reobserved
because of its peculiar blue index $B-R = -0.6$ in USNOC-A2.0. 
(2) carbon star identified by Cruz et al. (\cite{cruz03}). 
A first spectrum is shown in this work. (3) this object is a S-type evolved object}
\label{table01}
\end{table*}

\section{Results}

\subsection{Properties of the sample}

  Table\,\ref{table01} lists the main parameters  of the 25  objects that are under
analysis in this work. In the first column, the object serial number is given, 
following the numbering of Table 1 of Mauron et al.\,(\cite{mauron05}). 
Coordinates	$\alpha$  and $\delta$ (J2000) are given in the object 2MASS\,J
names as HHMMSS.ss$\pm$DDMMSS.s. The quantities $l$, $b$ are galactic coordinates 
in degrees. The $B$ \& $R$ magnitudes  are from the USNO-A2.0 catalogue
(Monet et al. \cite{monet98}) as given in the  2MASS catalogue. If these data are not 
available, we have adopted the  $B$ and $R$ values given in the APM database 
(Irwin \cite{irwin00}), or we 
have estimated them from the USNO-B1.0 data (Monet et al. \cite{monet03}). 
Uncertainties on $B$ and $R$ of the order of 0.4 mag (1\,$\sigma$). The  $J H K_{\rm s}$
magnitudes and the $J$$-$$K_{\rm s}$ colour are from the  2MASS catalogue, with 
uncertainties on $JHK_{\rm s}$ of  about 0.02-0.03 mag. in most cases.

One can see in Table\,\ref{table01} that the $R$ magnitudes are in the range $\sim 11$ up 
to $\sim 17$. Nine sources are brighter than $R=13.0$ and three are fainter 
than 16.0. The $B$\,$-$\,$R$ colour is often greater than 3.0\,. More precisely, of 
the 25 objects, 20  have their $B$\,$-$\,$R$ greater than 3.0. Here,  it is  interesting 
to note that if $B$\,$-$\,$R$ had been requested to be larger than 3.0 in our selection 
process, we would have missed the two most interesting objects (\#59 and \#81), 
with the faintest $K_{\rm s}$ magnitude and the largest distances (see below).
 
\begin{figure}[!ht]
\resizebox{8.5cm}{!}{\rotatebox{-90}{\includegraphics{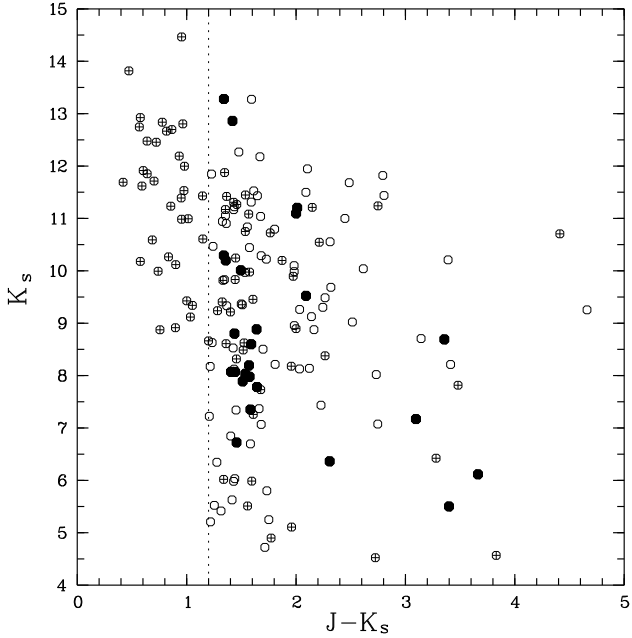}}}

\caption[]{Colour-magnitude diagram with $K_{\rm s}$ plotted 
versus $J$\,$-$\,$K_{\rm s}$ for known cool
halo C stars. Filled circles are objects of this work. Encircled crosses are the 
objects listed in TI98. Empty circles are objects found in other 
previous studies, 
with most of them being from Mauron et al. (\cite{mauron04}, \cite{mauron05}). 
The vertical dotted line at $J$-$K_{\rm s}$\,=\,1.2 approximatively separate
the CH-type stars to the left and the cool N-type stars to the right.}
\label{figure01}
\end{figure}

The listed  $K_{\rm s}$ magnitudes cover the range $\sim$\,6 to $\sim$\,13.
All  objects have $J-K_{\rm s}$\,$>$\,$1.30$.
Fig.\,\ref{figure01} shows the colour-magnitude diagram of our sample. The objects of
Table\,\ref{table01}  are plotted with filled circles. The encircled crosses
show the 77 objects listed in TI98, and the empty circles indicate
objects found in previous works aimed at discovering new faint and cool halo C stars
(e.g., Mauron et al. \cite{mauron04}, \cite{mauron05}; 
 Liebert et al. \cite{liebert00}).  The vertical line at 
$J$\,$-$\,$K_{\rm s}$\,$=$\,1.20 is the approximate limit between CH-type stars at 
left and cool AGB C stars at right. Only cool N-type objects were 
searched in the present study. In this diagram, a cool C star located at a 
distance of 50 kpc like the Large Magellanic Cloud (hereafter LMC)  has  
a typical magnitude  $K_{\rm s} \approx 10.8$.
One can note in Fig.\,\ref{figure01} that the number of objects in the 
upper right region $J-K_{\rm s} > 1.2$, $K_{\rm s} \ga 11$ has progressively
increased, with in particular  two objects, \#59 and \#83,
which have  $K_{\rm s}$\,=\,13.28 and $K_{\rm s}$\,=\,12.86  respectively.

There are two objects that were already known as carbon stars. One is \#63, 
that was already in  TI98  
under the name C*30 or APM\,0207-0211. 
It was reexamined by us  because of its intriguing colour   
$B$\,$-$\,$R$\,$=$\,$-0.6$ in 
the USNO-A2.0 catalog. Usually, this colour for cool C stars covers the range
from 2.5 to $\sim$\,4, or even redder. We wished to detect a possible
blue excess near 4500\,\AA\,, but found none. Our spectrum is very similar to  
that of TI98, the only difference being the presence  of a small H$\alpha$  
emission at 6563\,\AA\,. Because the APM catalogue gives  $R=12.16$ 
and  $B_J - R = 4.08$ (adopted in Table\,\ref{table01}), our conclusion is that 
the USNO-A2.0 data is  probably simply flawed. 

The other object which was known as a C star 
is \#69. This object was found by Cruz et al. (\cite{cruz03}) during their 
search for brown dwarfs. It was observed by us to obtain a first spectrum  
and because it was faint, $K_{\rm s}$\,$=$\,$11.2$, implying a large distance.

\subsection{Spectra}

\begin{figure}[!ht]
\resizebox{8cm}{!}{\rotatebox{-90}{\includegraphics{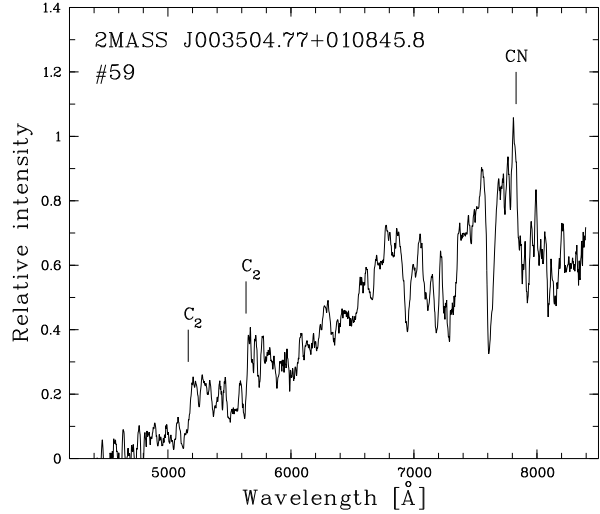}}}

\caption[]{ Spectrum of object \#59,  
the faintest source in the $K_{\rm s}$ band ($K_{\rm s}$\,=\,13.28, $R$\,$=$\,16.7). 
The spectrum is typical of cool N-type stars, with a flux increasing
with wavelength, and with characteristic molecular features, i.e. C$_{2}$ 
bands at $\lambda < 6000$\AA\, and the flux break at 
$\lambda \approx$\,7800\,\AA\, due to CN.}
\label{figure02}
\end{figure}

All  obtained spectra  are presented in the on-line Appendix. In Fig.\,\ref{figure02}, we
show only the spectrum of the most distant star (\#59). This
spectrum has the lowest S/N in our sample, but  the characteristics of 
a cool C star, such as a rising continuum and the C$_2$ and CN bands, are 
obvious. One object, \#80, is a S-type star, with strong VO 
absorption near 7400\,\AA\,. Its H$\alpha$ (6563\,\AA\,) line is in emission  
and this star displays important light variations (see below). Of the 25 C 
stars observed, 10 show the H$\alpha$ line 
in emission, representing 40\% of the sample, which is in agreement with 
previous findings of Mauron et al. (\cite{mauron04}, \cite{mauron05}).

%

\subsection{ Variability}

\begin{figure}[!ht]
\resizebox{8cm}{!}{\rotatebox{-90}{\includegraphics{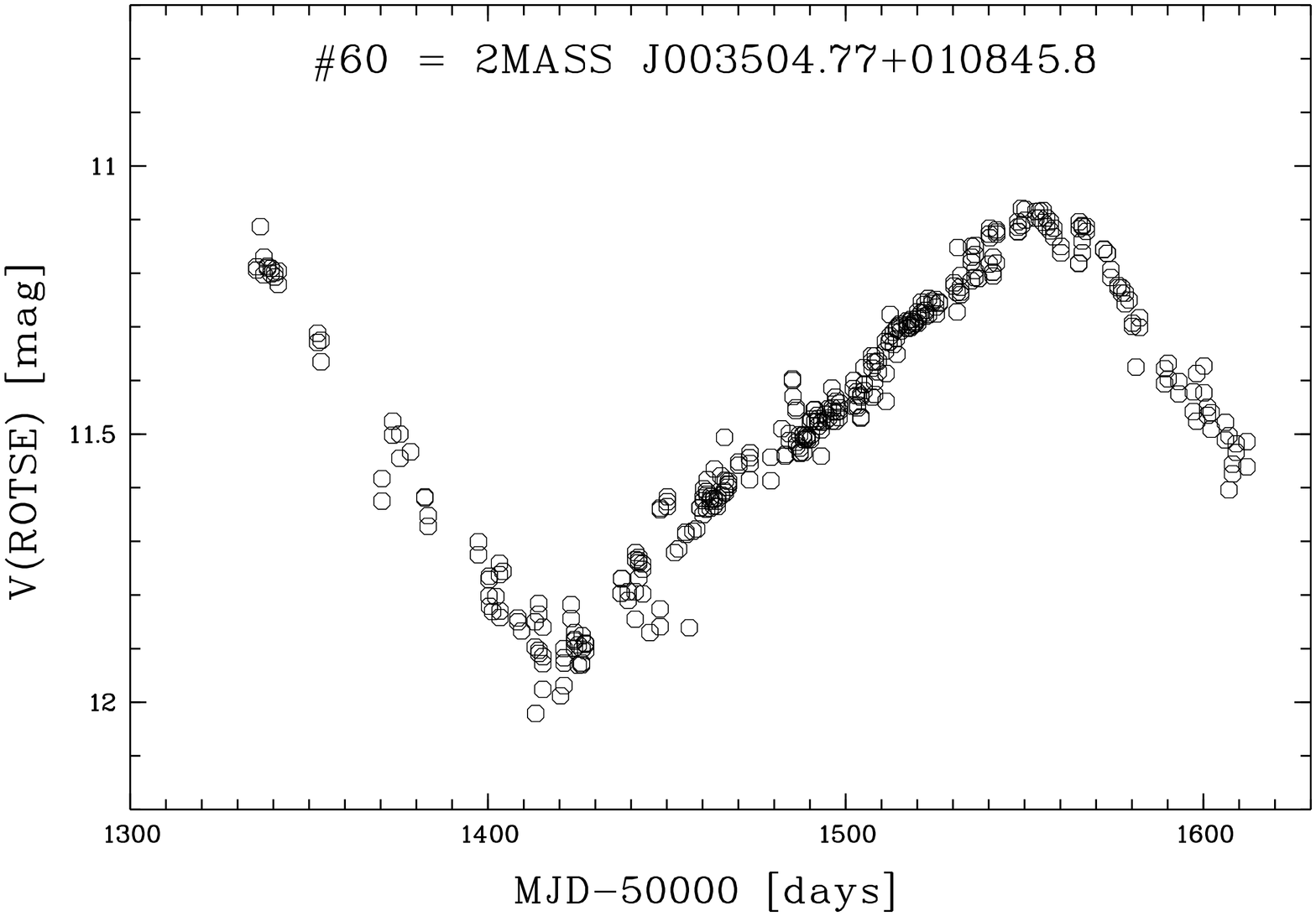}}}
\resizebox{8cm}{!}{\rotatebox{-90}{\includegraphics{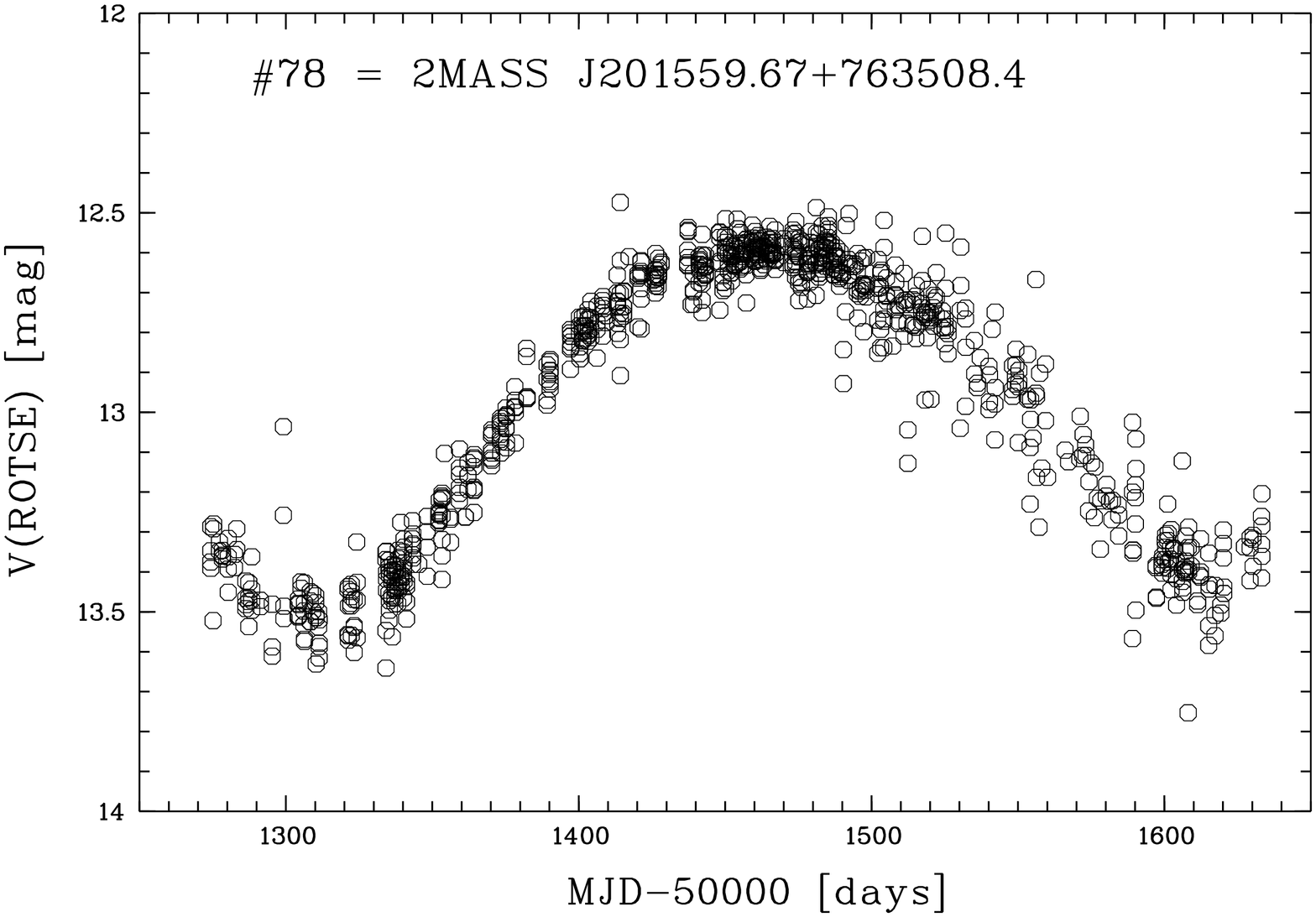}}}
\caption[]{Light curves for objects \#60 and \#78, from NSVS data. Only
the general trend should be considered and a few outliers have to be ignored.}
\label{figure03}
\end{figure}

In order to study variability of our objects, we have considered the Northern Sky 
 Variability Survey (NSVS) of Wo\'{z}niak et al. (\cite{wozniak04}). This survey 
provides light curves for sources with  $V$-band magnitudes of 8 to $\sim$15.5. 
The length of each record is typically 300 days. Some data points are obvious
outliers and should be ignored. Only the general trend of the light curves  must be
considered. Not all the objects
in our sample have a counterpart in the NSVS, especially the faintest ones.
In all cases in which data is available and sufficiently numerous,
they suggest regular or irregular variations with $V$-band amplitudes of  
about 1 mag. up to several magnitudes. 
For illustration, Fig.\,\ref{figure03} shows the light curves of  objects \#60 and \#78.  
Light curves for objects \# 66, 71, 73,  75, 76, 78, 79, 80, 82, and 83
are given in the on-line appendix. Note that the S star (\# 80) 
displays the largest variation and linearly decreases from $V$=12.5 to $V$=15.0 
over a period of 200 days.

 \section{Very red carbon stars in the halo}
 
     One of the result of this paper  is the discovery 
  of four very red  C stars  
  seen in Fig.\,\ref{figure01} as  filled  circles  with $J$\,$-$\,$K_{\rm s}$ between 3 and 4. 
  By searching the literature about high latitude, cool and faint C stars,
   we have found 8 other   objects out of the Galactic  plane
  having  a similarly red  $J$\,$-$\,$K_{\rm s}$ colour, $K_{\rm s}$\,$\ga$\,5.5,
  and $|b|$\,$>$\,20\degr.  In our research for these stars, we also 
  require the IRAS 12\,$\mu$m\, flux to be less than
  $\sim 7$\,Jy in order to exclude   nearby,
  very dusty  objects like IRC+10216 or AFGL 3068. 
  The  data for this sample of  12 cases are listed in Table\,\ref{table02}.
  Object \#82 has no 12\,$\mu$m flux because IRAS scans do not pass over a
  narrow lane in which this source lies. For 2MASS\,J194850.65-305831.7, 
  the 12\,$\mu$m data is a  conservative upper limit derived from the 
  examination of the IRAS Faint Source catalogue sources close to it.
      Distances to the Sun  and from the Galactic plane were derived 
  as explained below. One can see that the stars \#79 and \#82 are
  moderatly out of the plane by 1.3\,kpc, but all others are  clearly at  
  very large heights, and this remains true even if our distances are 
  overestimated by  $\sim$ 40\% (-2$\sigma$, see next section).
  
   The best studied case  is  IRAS\,12560+1656, discovered by Beichman 
   et al. (\cite{beichman90}) in a sample of faint IRAS sources located
  at high $b$ and having faint counterparts in the POSS plates.
  This object has been further investigated by 
  Groenewegen et al. (\cite{groe97}) who found 
  a remarkably low expansion velocity of its circumstellar envelope, measured 
  in CO (3\,\,km\,s$^{-1}$). They also found that this object is
  deficient in oxygen relative to the Sun, by $\sim$ 0.7\,\,dex. 
  IRAS\,08546+1732, not detected in CO,  is deficient as well.
  Therefore, our 12 objects deserve interest because it is established
  that already two of them are  metal-poor, with one having very unusual 
  wind properties.
  Because these 12 objects do not belong to the Galactic  disc, they may
  come from the debris of the Sgr dwarf and have a metal abundance 
  [Fe/H] around -1.0 (Van den Bergh \cite{vandenbergh00}).

\begin{figure}[!ht]
\resizebox{7.0cm}{!}{\rotatebox{-90}{\includegraphics{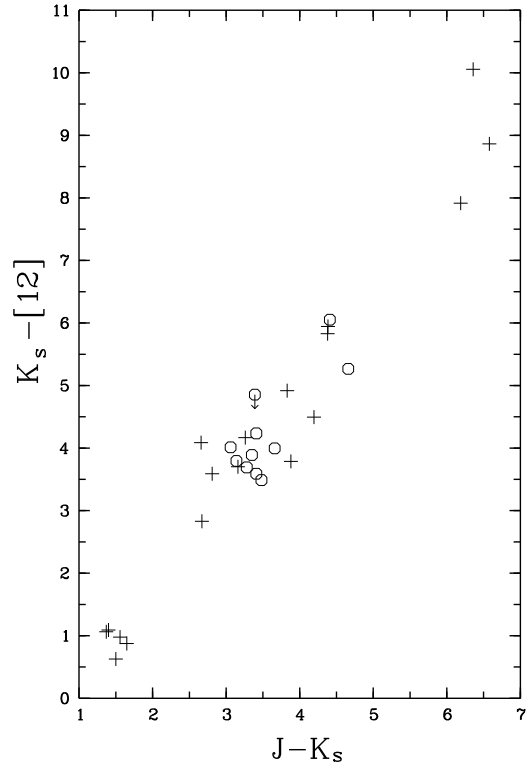}}}
\caption[]{Colour colour diagram $K_{\rm s}-[12]$ versus $J-K_{\rm s}$. Crosses represent
 a subsample of bright carbon stars of Groenewegen et al. (\cite{groe92}). Circles 
 indicate the objects of Table\,\ref{table02}, which appear to follow the trend of
 the bright carbon stars }
\label{figure04}
\end{figure}

   Are their infrared colours  different from those of C stars having 
   a much higher 12\,$\mu$m flux ? In order to test that, we considered the
  sample of 330 bright infrared C stars of Groenewegen 
  et al. (\cite{groe02}). Most of these stars are close to the Galactic  plane
  and might suffer some interstellar extinction in $J$ and $K_{\rm s}$. So we
  selected a subsample of their stars with $|b| > 10$\degr.  
  Their $f_{12}$ fluxes are between 13 and 250\,Jy. Figure\,\ref{figure04} shows
  the colour-colour plot with  $K_{\rm s}-[12]$ versus $J-K_{\rm s}$. 
  The [12] magnitude is taken to be zero for Vega, thus 
  [12]\,$=$\,$-$2.5\,log$_{10}$($f_{12}$/41.6), with $f_{12}$ in Jy. This plot
  shows that the objects in Table\,\ref{table02} are located on the sequence of ordinary
  disc C stars going from
  warm cases  at lower left to very cool dusty objects at upper right.
   The 12 objects of Table\,\ref{table02}  do not seem  different from brighter objects
   as far as the $J$-$K_{\rm s}$ and $K_{\rm s}$-[12] colours are concerned. 
     
      To conclude, in addition
   to IRAS\,12560+1656 and IRAS\,08546+1732 quoted above,
  the 10 other stars of Table\,\ref{table02}  may be useful to study  mass loss of 
  low metallicity AGB C  stars. They  are generally much closer to 
  the Sun than similar carbon stars in the Magellanic Clouds. 
   
 \begin{table*}[!ht]
        \caption[]{List of known very red C stars with $J-K_{\rm s} >3$, 
	$f_{12} < 7 $ \,Jy,
	 and located out of the Galactic  plane.
	After the names and galactic coordinates, this table provides the 
	IRAS 12\,$\mu$m flux, $K_{\rm s}$ and $J-K_{\rm s}$ from 2MASS, 
	distances $d$ and heights $Z$ above the Galactic  plane.}
        \begin{center}
        \begin{tabular}{ccrrrrrrrl}
        \noalign{\smallskip}
        \hline
        \hline
        \noalign{\smallskip}
  IRAS name & 2MASS  name & $l$\hspace{5mm}  & $b$\hspace{5mm} & $f_{12}$ & $K_{\rm s}$\,\,\,   & $J-K_{\rm s}$  & 
  $d$\hspace{2mm} &
  $Z$\,\,\,     & Note\\
            &             &  (deg)\,\,\,\, & (deg)\,\,\,\, &  (Jy)    & (mag) &  (mag) & (kpc)& (kpc) & -\\
        \noalign{\smallskip}
        \hline
        \noalign{\smallskip}

   01582$+$0931  & \object{2MASS J020056.14$+$094535.6}  & 149.845 &$-$49.446 & 2.27  &  7.17 & 3.06 &   7 & $-$5.5 &  \#62\\
   03242$+$1429  & \object{2MASS J032659.91$+$143956.9}  & 169.816 &$-$33.690 & 0.59  &  8.21 & 3.41 &  13 & $-$7.5 & 1\\
   03582$+$1819  & \object{2MASS J040109.72$+$182808.1}  & 173.496 &$-$25.252 & 1.06  &  9.25 & 4.66 &  17 & $-$7.5 & 1\\
   04188$+$0122  & \object{2MASS J042127.25$+$012913.4}  & 192.177 &$-$31.987 & 3.37  &  6.42 & 3.28 &   6 & $-$3.5 & 2\\
   08427$+$0338  & \object{2MASS J084522.27$+$032711.2}  & 223.486 &$+$26.816 & 6.50  &  6.25 & 3.41 &   6 & $+$2.5 & 3\\
   08546$+$1732  & \object{2MASS J085725.82$+$172051.9}  & 210.261 &$+$35.437 & 0.57  & 10.71 & 4.41 &  36 & $+$21.0& 4\\
   12560$+$1656  & \object{2MASS J125833.50$+$164012.2}  & 312.253 &$+$79.413 & 0.77  &  7.82 & 3.48 &  11 & $+$11.0& 5\\
    not IRAS     & \object{2MASS J194850.65$-$305831.7}  &   9.433 &$-$25.076 & $< 0.3$& 10.21& 3.39 &  25 & $-$10.0& 6\\
   20005$+$7737  & \object{2MASS J195840.16$+$774526.2}  & 110.190 &$+$23.019 & 0.50  &  8.69 & 3.35 &  14 & $+$5.5 &  \#77\\
   20176$-$1458  & \object{2MASS J202027.66$-$144927.2}  & 029.047 &$-$26.265 & 0.45  &  8.71 & 3.14 &  14 & $-$6.0 & 6\\
   21064$+$7749  & \object{2MASS J210522.23$+$780116.2}  & 112.308 &$+$20.130 & 5.88  &  6.12 & 3.66 &   4 & $+$1.3 & \#79\\
    not IRAS     & \object{2MASS J215526.97$+$234214.4}  &  78.911 &$-$23.760 &  -    &  5.50 & 3.40 &   3 & $-$1.3 & \#82\\
      
    \noalign{\smallskip}
   \hline
\end{tabular}
\end{center}
{\small Notes:  (1) discovered by Liebert et al. (\cite{liebert00}).\,\, (2) 
 discovered by TI98 as APM 0418+0122.\,\, (3) this star is CGCS\,6306 in the Catalogue of
 Galactic carbon stars, Alksnis et al. (\cite{alksnis01}).\,\, (4) discovered by 
 Cutri et al. (\cite{cutri89}).
 \,\, (5) discovered by Beichman et al. (\cite{beichman90}).\,\, (6) discovered by  Mauron
 et al. (\cite{mauron04}).} 
\label{table02}
\end{table*}
  
\section{Distances and location in the halo}

 The distances of our objects were estimated by using their  
 observed $J$-$K_{\rm s}$  colour and $K_{\rm s}$ magnitude.
 The $K_{\rm s}$ band is used because it is less sensitive to temporal variations 
 than other filters and suffers less interstellar  extinction. Absolute magnitudes 
 were determined as in Mauron et al. (\cite{mauron04}, \cite{mauron05}).
 Briefly,  $M_{\rm K_s}$ values are obtained by considering the 
 $K_{\rm s}$ magnitudes of LMC carbon stars averaged over colour bins of 0.1\,mag 
 and adding 0.5 mag. The colour-magnitude diagram $K_{\rm s}$ versus $J-K_{\rm s}$ 
 of the LMC can be seen in Fig.\,3 of Nikolaev and Weinberg (\cite{niko00}). The 
 supplementary 0.5-mag term is due to the average
 difference between C stars in the LMC and C stars in the Sgr dwarf
 galaxy (see Mauron et al. \cite{mauron04} for details). This calibration on the
 C stars of Sgr is adopted because a majority of halo C stars originate
 from this dwarf galaxy. If they were calibrated on LMC C stars, our
 objects would have distances and $Z$ values $\sim$ 25\% larger. 
 The scatter on the adopted absolute magnitude $M_{\rm K_s}$ is around
  0.3-0.4 (1$\sigma$). Concerning the S star (\# 80), it was assumed 
  that its $M_{\rm K_s}$ was similar to that of C stars.
  
  For each star,  interstellar colour excesses 
 $E_{\rm B-V}$ were taken from the maps of Schlegel 
 et al. (\cite{schlegel98}) and a colour excess of 0.13 was adopted for LMC stars. 
 The $J$ and $K_{\rm s}$ extinctions were calculated as 
 A$_{\rm J}$\,=\,0.902\,$E_{\rm B-V}$ and  A$_{\rm K_s}$\,=\,0.367\,$E_{\rm B-V}$. 
 In general, these corrections for interstellar reddening are  low, 
 but for a few stars $E_{\rm B-V}$ is of the order of 0.5-0.6 mag, 
 so that A$_{\rm K_s}$ reaches $0.2$ mag and
 cannot be neglected. Table\,\ref{table03}  shows the adopted values of
  $E_{\rm B-V}$, the dereddened colour ($J-K_{\rm s})$$_{0}$ , the
  absolute magnitude  $M_{\rm K_s}$ and the derived 
  distance $d$ in kpc.   Due to the scatter on $M_{\rm K_s}$, the
  uncertainty on distances is about 20 percent (1$\sigma$).

  It can be seen in Table\,\ref{table03} that distances are 
  from $\sim 5$\,kpc to $\sim 110$\,\,kpc. Two stars have $d$\,$<$\,5\,\,kpc,
  15 have $d$ between 5 and 20\,\,kpc and 8 have $d$ between 20 and 110\,\,kpc.
  Some of these distances could be improved in the future if the stars  are 
  shown to be periodic variables. For example, if we assume that stars \#60 and \#78
  (with light curves in Fig.\,\ref{figure03}) are periodic, we can apply the $K$-band 
  period-luminosity relation for
  semiregular variables established by  Knapp et al. (\cite{knapp03}). The
  NSVS light curves suggest that their periods could be  240 and 320 days,
  respectively. Then, with the P-$M_{\rm K}$ relation 
  $M_{\rm K} = -1.34 $\,log\,$ P -4.5 $, and assuming  that 
  the (average) $K$ magnitude is not too different from $K_{\rm s}$, we find
  for \#60, $M_{\rm K}$\,$=$\,$-7.69$, $d$\,$=$\,7.6\,kpc, a distance close to
  the one in Table 3 (5.9 kpc). As for \#78, we obtain
  $M_{\rm K}$\,$=$\,$-7.86$, $d$\,$=$\,14.7\,kpc, and this distance is 
  40\% larger than that found in Table\,\ref{table02}  (10.5\,kpc).

   \begin{figure*}[!ht]
    \resizebox{\hsize}{!}{
    {\rotatebox{-90}{\includegraphics{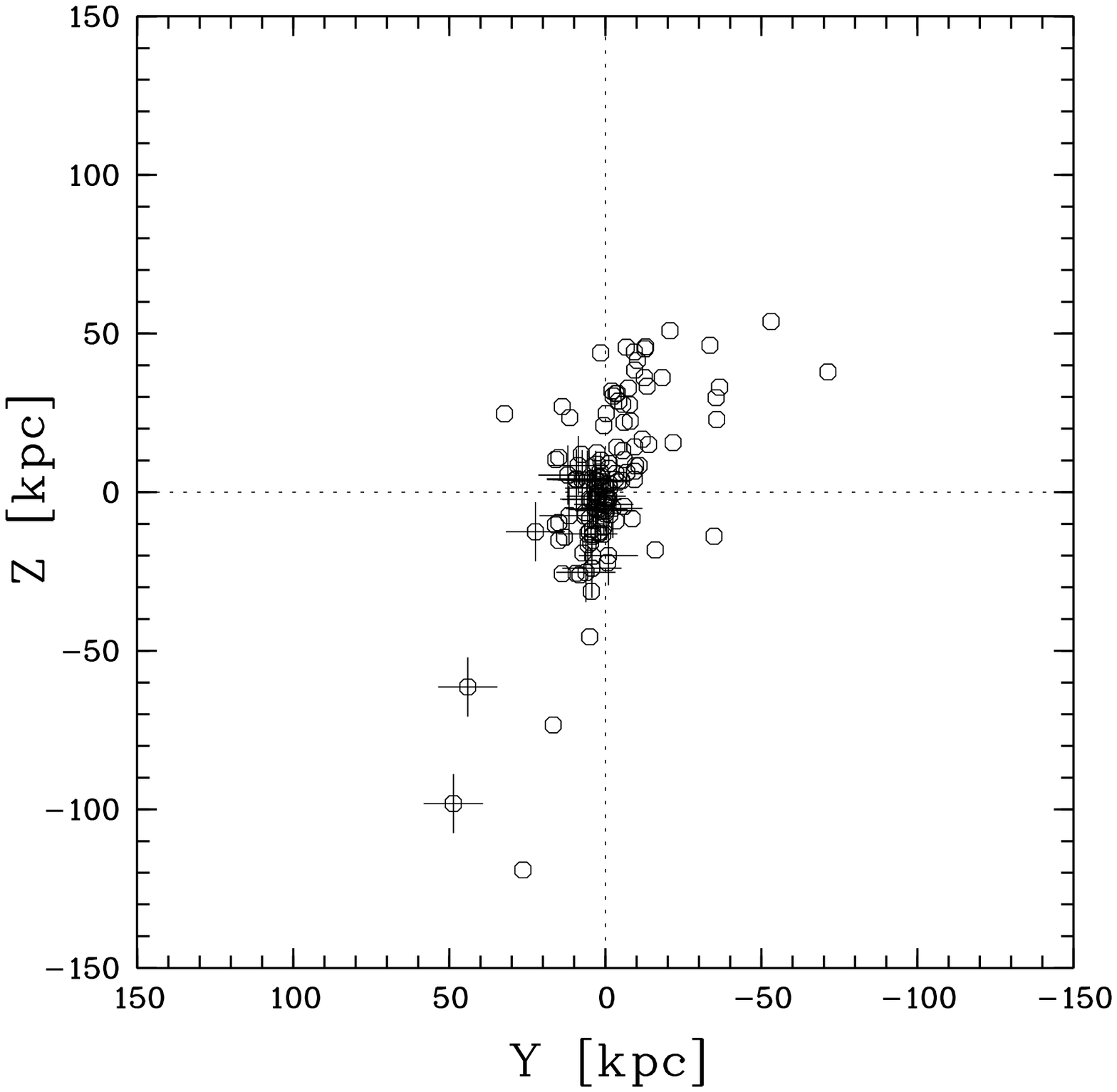}}}
    {\rotatebox{-90}{\includegraphics{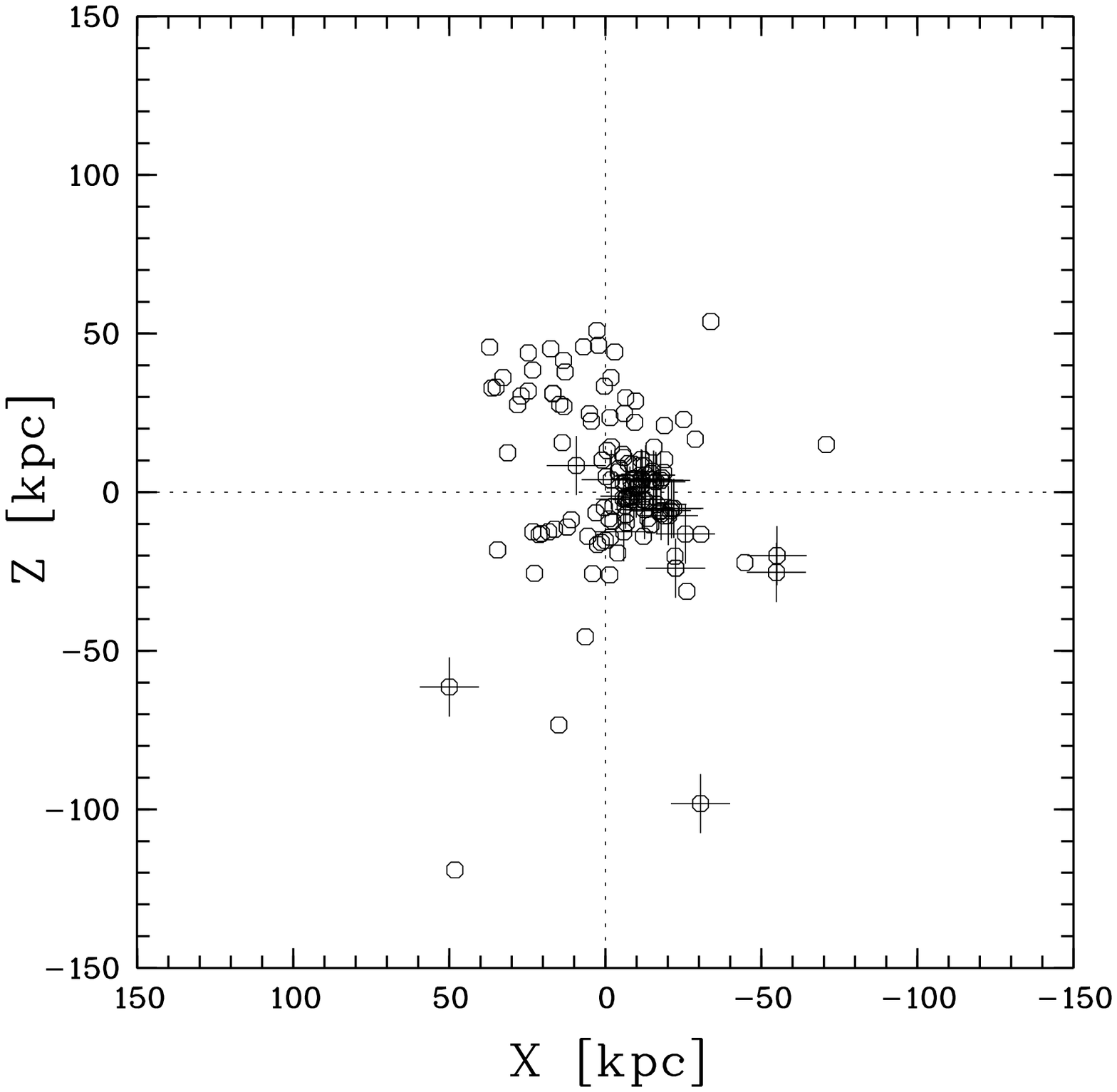}}} }

     \caption[]{ Plots in galactocentric $XYZ$ of the observed   halo C stars, with those 
     found in this paper indicated by circle with a  overplotted cross. {\it Left panel}:  
     in this $Z$ versus $Y$ the   Sgr Stream is seen nearly edge-on, slightly tilted
     clockwise with respect to the $Z$ axis. 
     Note that $Y$ is positive to the left.  The two new very distant C stars
      found in this paper(\# 59 and \#81) are on the lower left. Their
      distances to the main plane of the Sgr Stream are about 25 kpc.
      {\it Right panel}: $Z$ versus $X$ plot. $X$ is positive to the left 
      and $X_{\odot}$\,$= -$\,8.5 kpc. The positions of the C stars at $Z < -50$\,kpc 
      suggest that the Sgr debris have a broad extent, $\sim$ 100 kpc, 
      along the X-coordinates.}

     \label{figure05}
     \end{figure*}

Galactocentric $XYZ$ coordinates were calculated in the system of 
Newberg et al. (\cite{newberg03}), in which the Sun is at $X$\,$=$\,$-8.5$\,kpc, 
$Y$\,$=$\,$0$, $Z$\,$=$\,$0$; the $Y$-axis is positive towards  
$l$\,$=$\,$+90$$\degr$ and  the $Z$ axis is positive to $b$\,$=$\,$+90$$\degr$.
Figure\,\ref{figure05} displays the $YZ$ and $XZ$ plots of previously known C stars and
those of Table\,\ref{table01}. In this system, the  Sgr Stream as traced with the M-type giants 
(Majewski et al. \cite{majewski03}) is approximatively given by the following equation  
(Newberg et al. \cite{newberg03}):

$-0.064$\,$X$\,$+$\,$0.970$\,$Y$\,$+$\,0.233\,$Z$ $\,$+$\,$0.232$\,$\,$=$\,\,$0$

In Table\,\ref{table03}, last column, the distance $D$ of each object to this average plane
is derived. We chose $D$ to have the sign of $Y$. It can be seen  that $D$ is 
relatively small when compared with the range of
distances. More specifically, out of 25 stars, there are 14 with $D < 5$\,kpc.
Some stars are quite distant from the Sun, like \#65 at 53\,kpc and \#69 at 51 kpc,
but their $D$ is very small, $\sim 4$ and $-2$\,kpc. This suggests that they belong 
to the Sgr Stream, although supplementary observations, especially radial velocities, are
needed to confirm this.

There are two new stars that are especially distant, 
\#59 at $\sim$110\,kpc and \#81 at $\sim$95\,kpc. Figure\,\ref{figure05} shows that
four carbon stars are now known in the region $Z < -60$\,kpc and none are
known in the symetrical region $Z > +60$\,kpc, although a large number
have been discovered closer to us and with  $0 \la Z \la 50$\,kpc. 
This asymetry qualitatively supports the existence of an extended branch of 
the Sgr Stream as in the model of Law et al. (\cite{law05}) 
(see Fig.\,4 of Mauron et al. \cite{mauron05}).
The width of this branch along the $X$ direction is $\sim$\,90\,kpc, which
is in fair agreement with the positions of our four C stars.

We note however that, recently, the Sgr Stream has been
mapped by analysing the data of Sloan survey 
(Belokurov et al. \cite{belokurov06}, Fellhauer et al. \cite{fellhauer06}).
The map puts strong constraints on the Stream geometry and 
is well explained within a model  where there
is no matter at $Z < -60$\,kpc. It is then  possible that our four C stars with
high negative $Z$ are part of another stream. It will be interesting in the future
to see if any new distant star can be detected close to them. 
In both cases, the fact that cool carbon stars are found 
indicate that an intermediate-age population is present in this region of the halo.

\begin{table*}[ht]
        \caption[]{Properties of the halo C stars. The table lists the running number, the galactic
	coordinates $l$ and $b$ in degrees, the interstellar colour excess measured in the
	direction of the object, $E_{\rm B-V}$ from Schlegel et al. maps, the dereddened $J-K_{\rm s}$ colour,
	the adopted $K_{\rm s}$-band absolute magnitude, the distance $d$ to the Sun in kpc, the galactocentric
	$XYZ$ coordinates in kpc, and the distance $D$ to the Sgr Stream average plane, in kpc.}
        \begin{flushleft}
        \begin{tabular}{cccccrrrrrr}
        \noalign{\smallskip}
        \hline
        \hline
        \noalign{\smallskip}
No.& $l$ & $b$ & $E_{\rm B-V}$ & $(J-K_{\rm s})_0$ & $M_{\rm K_s}$ & $d$\,\,\,\, & $X$ & $Y$ & $Z$ & $D$\\
        \noalign{\smallskip}
        \hline
        \noalign{\smallskip}
	
 59 & 114.351 &$-$61.453 & 0.020 & 1.329 & $-$6.97 &111.8 & $-$30.5  & 48.7    &$-$98.2 &  26.5\\
 60 & 127.393 &$-$22.689 & 0.050 & 1.429 & $-$7.15 &  5.9 & $-$11.8  &  4.3    & $-$2.3 &   4.7\\
 61 & 135.175 &$-$24.325 & 0.053 & 1.611 & $-$7.42 & 18.0 & $-$20.2  & 11.6    & $-$7.4 &  11.0\\
 62 & 149.845 &$-$49.446 & 0.068 & 3.060 & $-$7.16 &  7.3 & $-$12.6  &  2.4    & $-$5.5 &   2.1\\
 63 & 163.123 &$-$58.546 & 0.032 & 1.478 & $-$7.25 & 28.1 & $-$22.5  &  4.3    &$-$24.0 &   0.2\\
 64 & 162.049 &$-$36.082 & 0.291 & 1.183 & $-$6.56 & 22.4 & $-$25.7  &  5.6    &$-$13.2 &   4.2\\
 65 & 172.347 &$-$28.475 & 0.316 & 1.832 & $-$7.64 & 53.1 & $-$54.8  &  6.2    &$-$25.3 &   3.9\\
 66 & 165.713 &$-$21.582 & 0.164 & 1.349 & $-$7.01 & 14.1 & $-$21.2  &  3.2    & $-$5.2 &   3.5\\
 67 & 181.911 &$-$31.637 & 0.289 & 1.415 & $-$7.13 & 11.1 & $-$17.9  & $-$0.3  & $-$5.8 &$-$0.3\\
 68 & 176.133 &$-$25.619 & 0.472 & 1.284 & $-$6.89 &  8.9 & $-$16.5  &  0.5    & $-$3.9 &   0.9\\
 69 & 181.245 &$-$23.296 & 0.587 & 1.697 & $-$7.53 & 50.6 & $-$55.0  & $-$1.0  &$-$20.0 &$-$1.9\\
 70 & 190.204 &$-$20.452 & 0.164 & 1.502 & $-$7.28 & 14.6 & $-$21.9  & $-$2.4  & $-$5.1 &$-$1.9\\
 71 & 140.861 &$+$22.738 & 0.177 & 1.417 & $-$7.13 &  9.8 & $-$15.5  &  5.7    &  3.8   &   7.7\\
 72 & 161.381 &$+$21.086 & 0.080 & 1.363 & $-$7.03 & 10.3 & $-$17.7  &  3.1    &  3.7   &   5.2\\
 73 & 169.337 &$+$22.917 & 0.071 & 1.545 & $-$7.33 &  8.5 & $-$16.2  &  1.5    &  3.3   &   3.5\\
 74 &  26.142 &$+$22.867 & 0.316 & 1.188 & $-$6.58 & 21.5 &   9.3    &  8.7    &  8.4   &  10.1\\
 75 &  48.195 &$+$21.356 & 0.082 & 1.399 & $-$7.10 & 10.7 &  $-$1.9  &  7.4    &  3.9   &   8.4\\
 76 & 107.672 &$+$23.332 & 0.093 & 1.593 & $-$7.39 & 10.7 & $-$11.5  &  9.3    &  4.2   &  11.0\\
 77 & 110.190 &$+$23.019 & 0.155 & 3.273 & $-$7.08 & 13.9 & $-$12.9  & 12.0    &  5.4   &  13.9\\
 78 & 109.469 &$+$21.659 & 0.218 & 1.458 & $-$7.21 & 10.5 & $-$11.8  &  9.2    &  3.9   &  10.8\\
 79 & 112.308 &$+$20.130 & 0.548 & 3.370 & $-$7.03 &  3.9 &  $-$9.9  &  3.4    &  1.3   &   4.5\\
 80 &  68.200 &$-$21.673 & 0.092 & 2.258 & $-$7.50 &  5.8 &  $-$6.5  &  5.0    & $-$2.2 &   5.0\\
 81 &  37.019 &$-$39.965 & 0.069 & 1.382 & $-$7.07 & 95.6 &  50.0    & 44.1    &$-$61.4 &  25.5\\
 82 &  78.911 &$-$23.760 & 0.066 & 3.363 & $-$7.04 &  3.2 &  $-$7.9  &  2.9    & $-$1.3 &   3.2\\
 83 &  83.298 &$-$28.940 & 0.058 & 2.130 & $-$7.56 & 25.8 &  $-$5.9  & 22.4    &$-$12.5 &  19.5\\

    \noalign{\smallskip}
   \hline
\end{tabular}
\end{flushleft}
\label{table03}
\end{table*}

\section{Conclusions} 

By searching the 2MASS catalogue for rare cool carbon stars located out of the
Galactic  plane ($|b| > 20$\degr), we have found 23 new cases with 
$K_{\rm s}$ magnitudes covering the
range $\sim$\,6 to 13.3. Their colours and spectra are typical
of AGB stars, which is also supported by important variability 
for those objects bright enough to be detected by the NSVS survey.

\smallskip

We have discovered four cases of C stars with $J$\,$-$\,$K_{\rm s}$\,$>$\,$3$ and located 
well above the Galactic  plane. After including those already published
in the literature, we could made a list of 12 objects with similar properties.
Previous works (Groenewegen et al. \cite{groe97}) show that two of them are 
deficient in oxygen and one has a very small wind expansion velocity. This
suggests that some (or all) members of this sample could be metal deficient as well.
Because they are closer to us than the Magellanic Clouds,  
supplementary observations of these stars, especially their CO emission, could help 
to better know  AGB winds at low metallicity. 

\smallskip

Distances were estimated by using the 2MASS $J$ and $K_{\rm s}$ data and 
by assuming our objects to be similar to those of
the Sagittarius dwarf galaxy. It is found that two objects 
are at remarkably large distances from the Sun ($\sim 100$\,kpc). Their galactocentric
$XYZ$ coordinates suggest that they might belong, together with two other
distant carbon stars previously found, to a distant loop
of the Sgr stream extending at large negative $Z$. Additional detections
of distant stars in this region of the halo should help to confirm this
conclusion.

\begin{acknowledgements}
We would like to thank  our referee, E.\,\,Lagadec, for useful comments. We thank also
 the staff of Observatoire de Haute-Provence, which is supported
by the French Centre National de Recherche Scientifique.
We  acknowledge the use of  the Two Micron All Sky Survey
(2MASS), which is a joint project of the University of Massachusetts and  the Infrared
Processing and Analysis Centre / California Institute of Technology, funded
by  the National Aeronautics and Space Administration (NASA) and the
 National Science Foundation (NSF). 
This publication also makes use of the data from the Northern Sky Variability 
Survey (NSVS) created jointly  by the Los Alamos National Laboratory and University of Michigan.
The NSVS was funded by the Department of Energy,  NASA and  NSF. Finally, this
work benefited from using the CDS database of Strasbourg (France).

\end{acknowledgements}



\Online

\appendix

\section{Appendix A}
 
\begin{figure*}
\caption[]{ Stars observed at Byurakan}
\resizebox{8.5cm}{!}{\rotatebox{-90}{\includegraphics{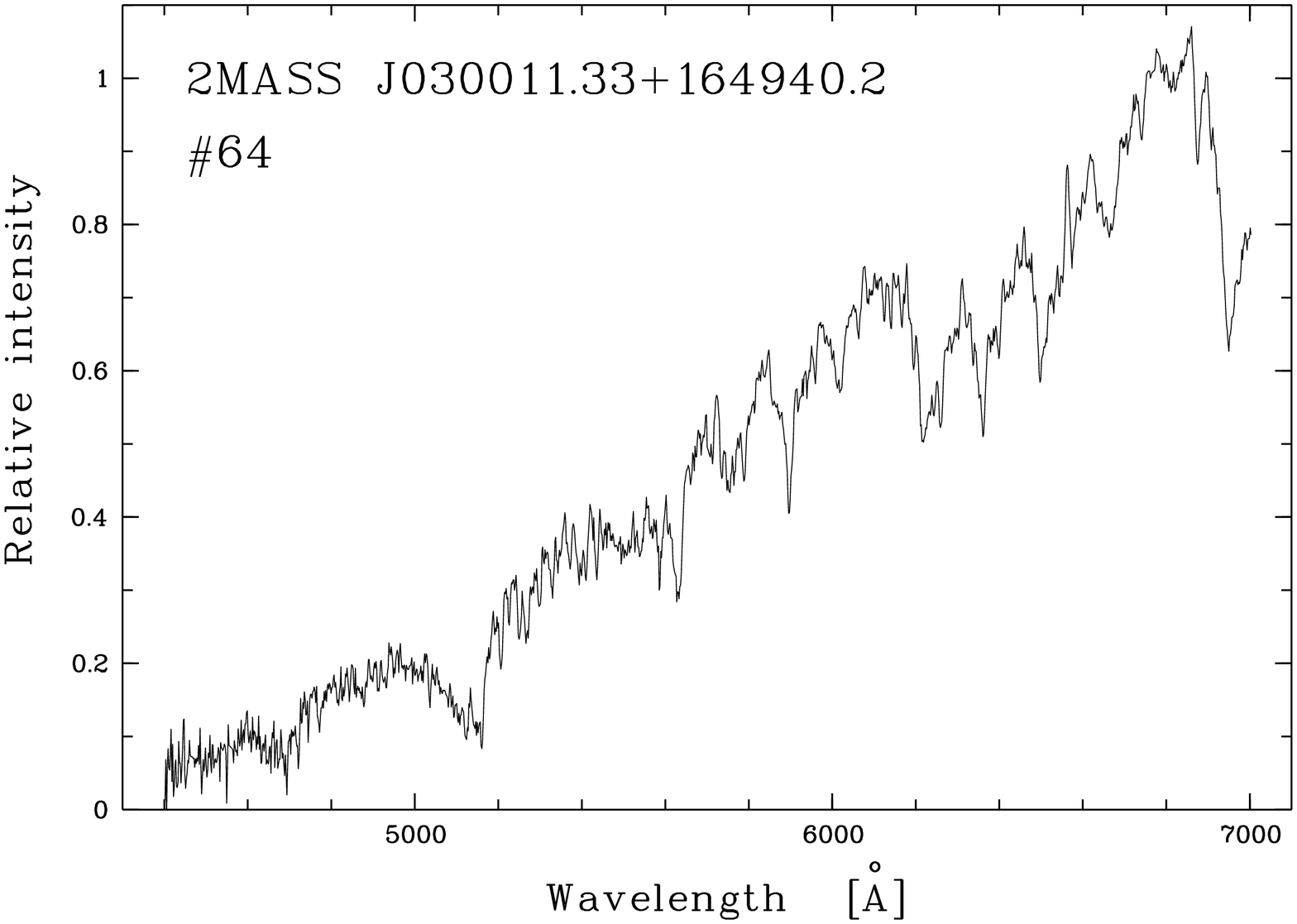}}}
\resizebox{8.5cm}{!}{\rotatebox{-90}{\includegraphics{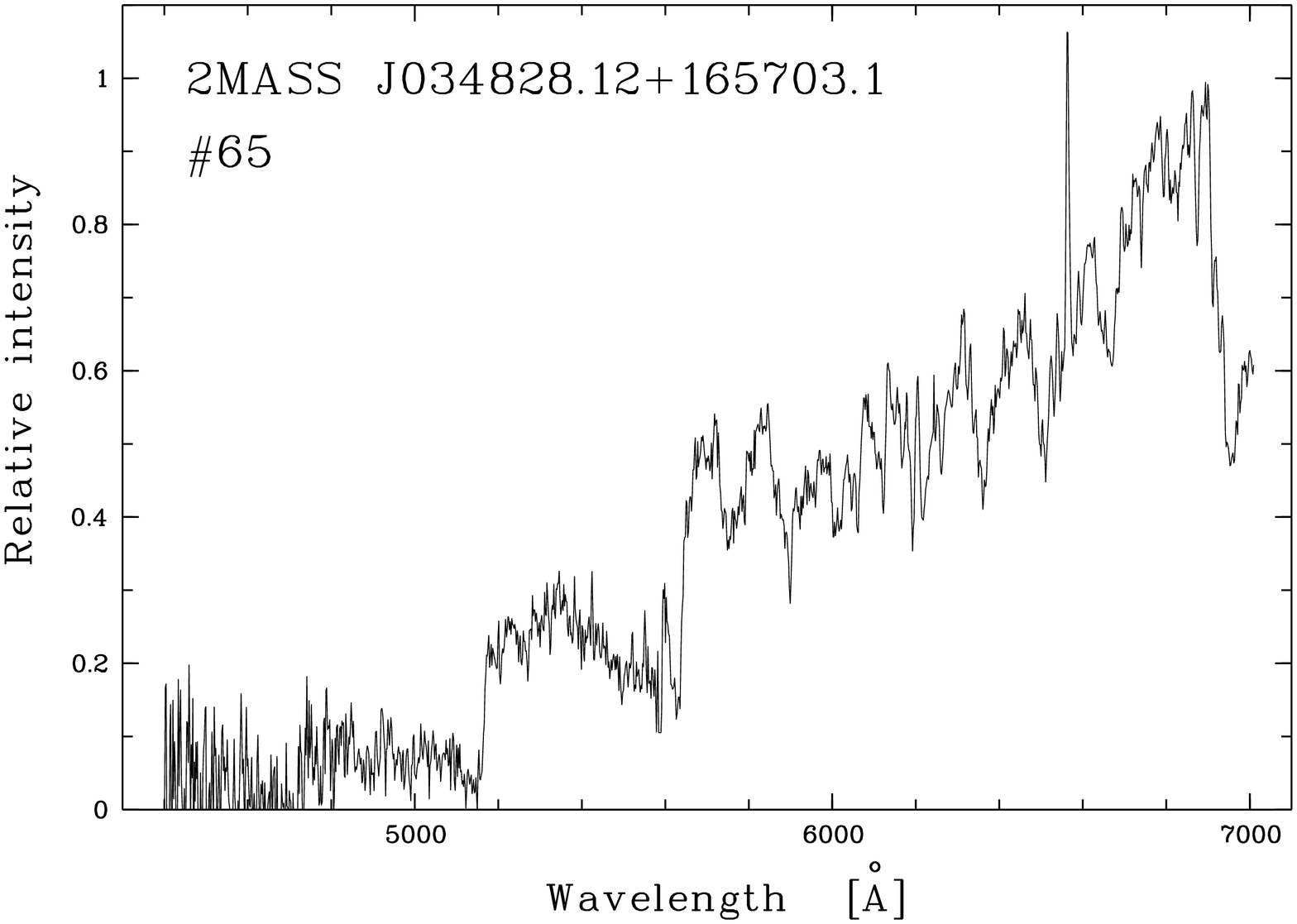}}}
\resizebox{8.5cm}{!}{\rotatebox{-90}{\includegraphics{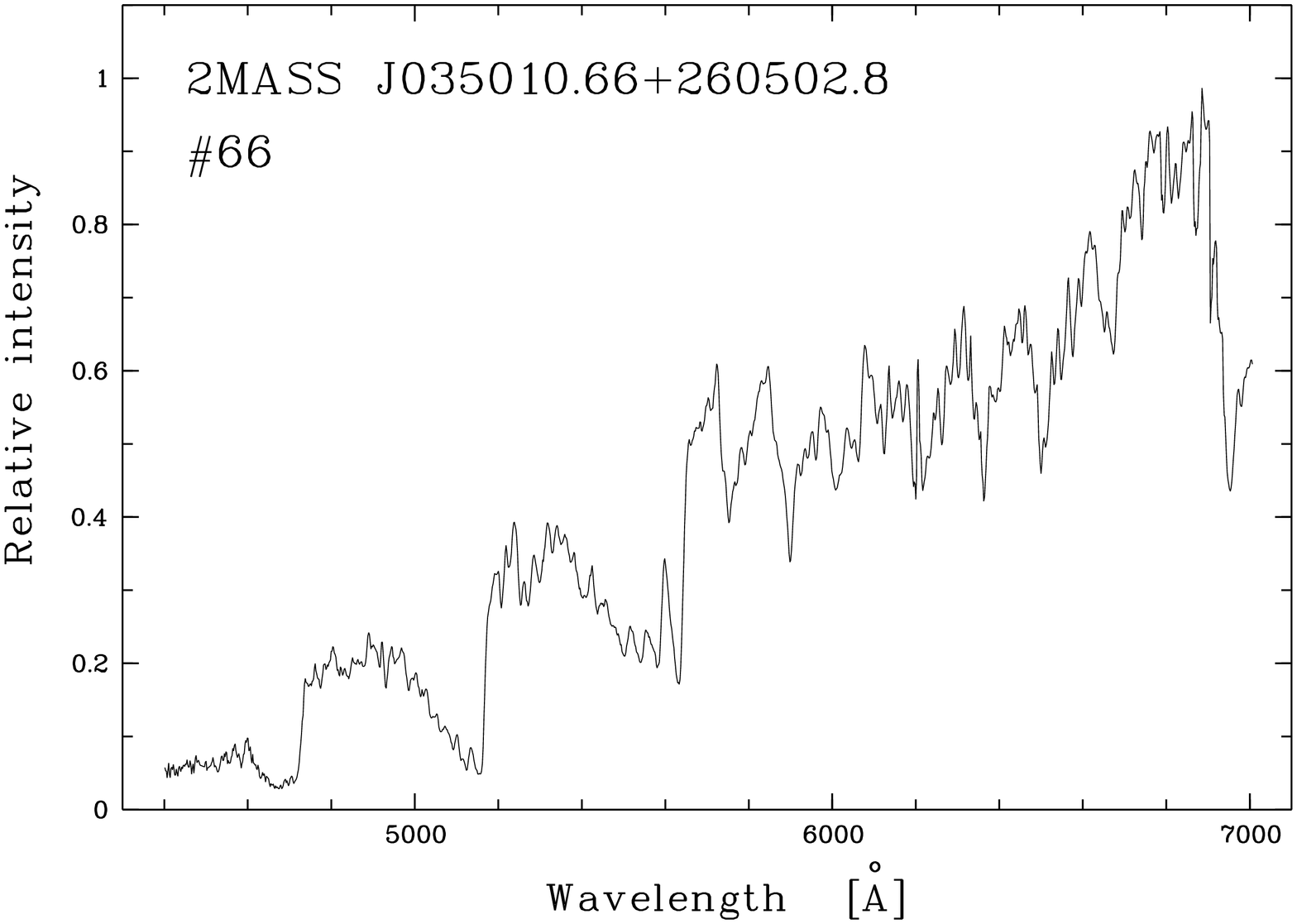}}}
\resizebox{8.5cm}{!}{\rotatebox{-90}{\includegraphics{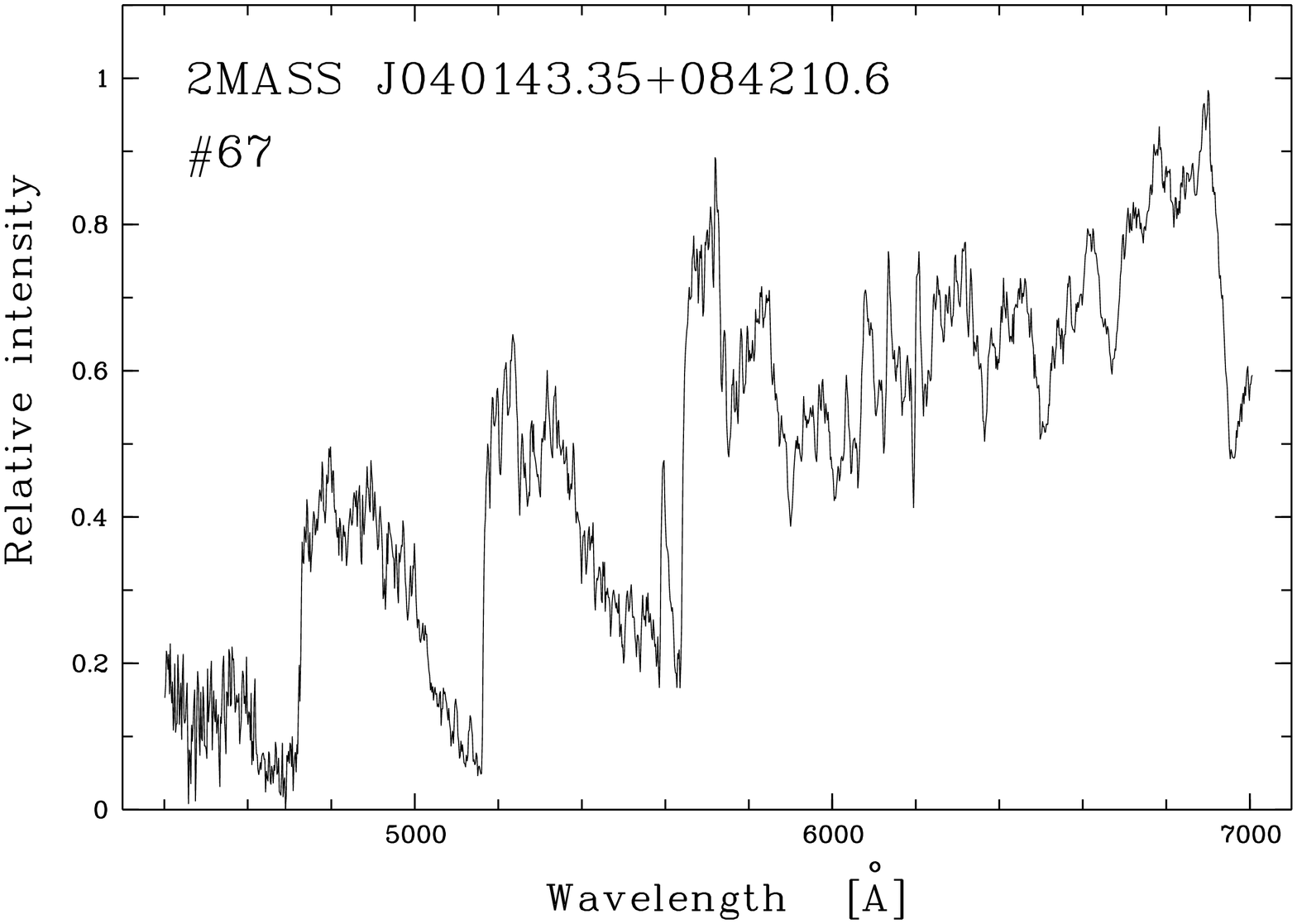}}}
\resizebox{8.5cm}{!}{\rotatebox{-90}{\includegraphics{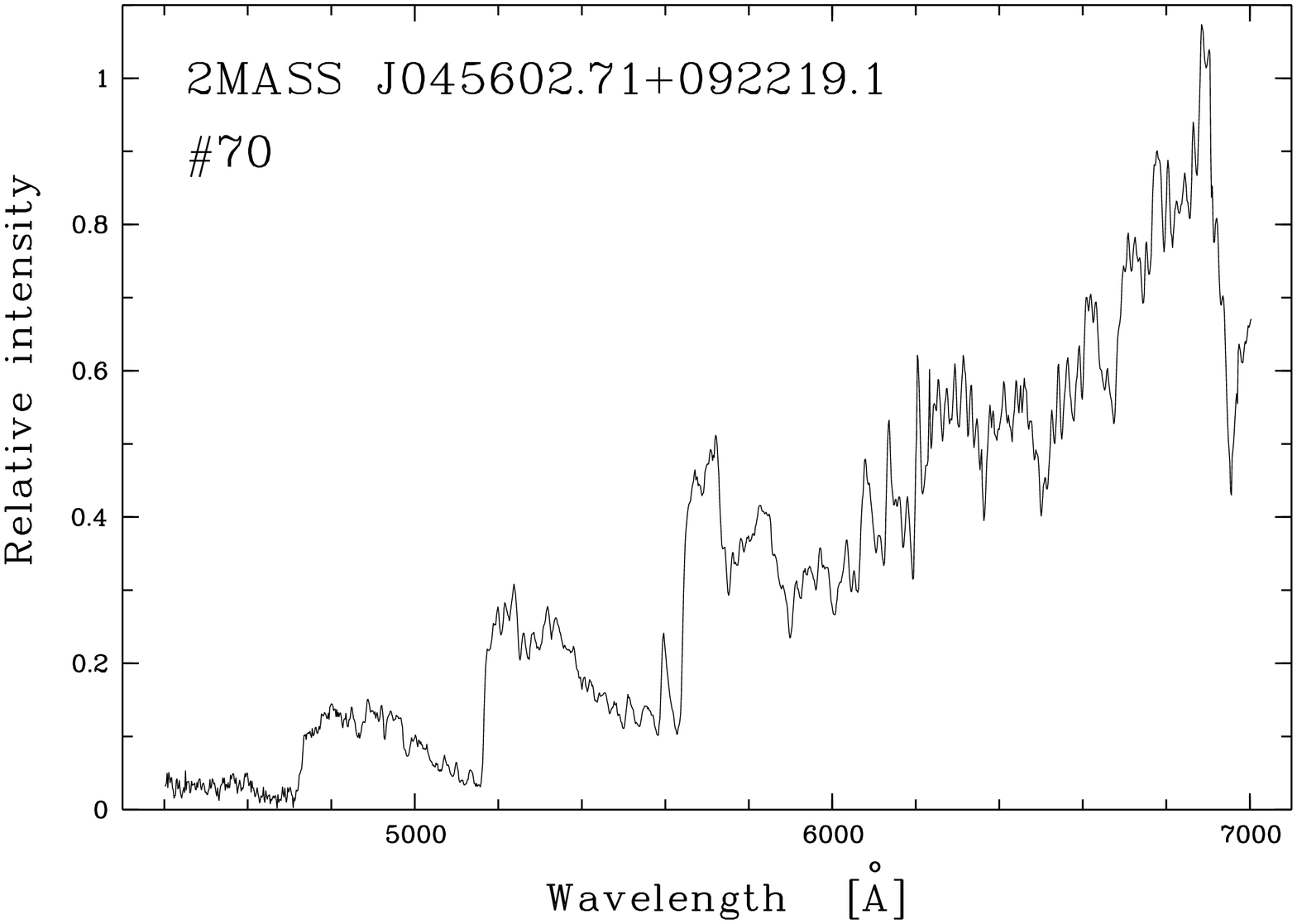}}}
\resizebox{8.5cm}{!}{\rotatebox{-90}{\includegraphics{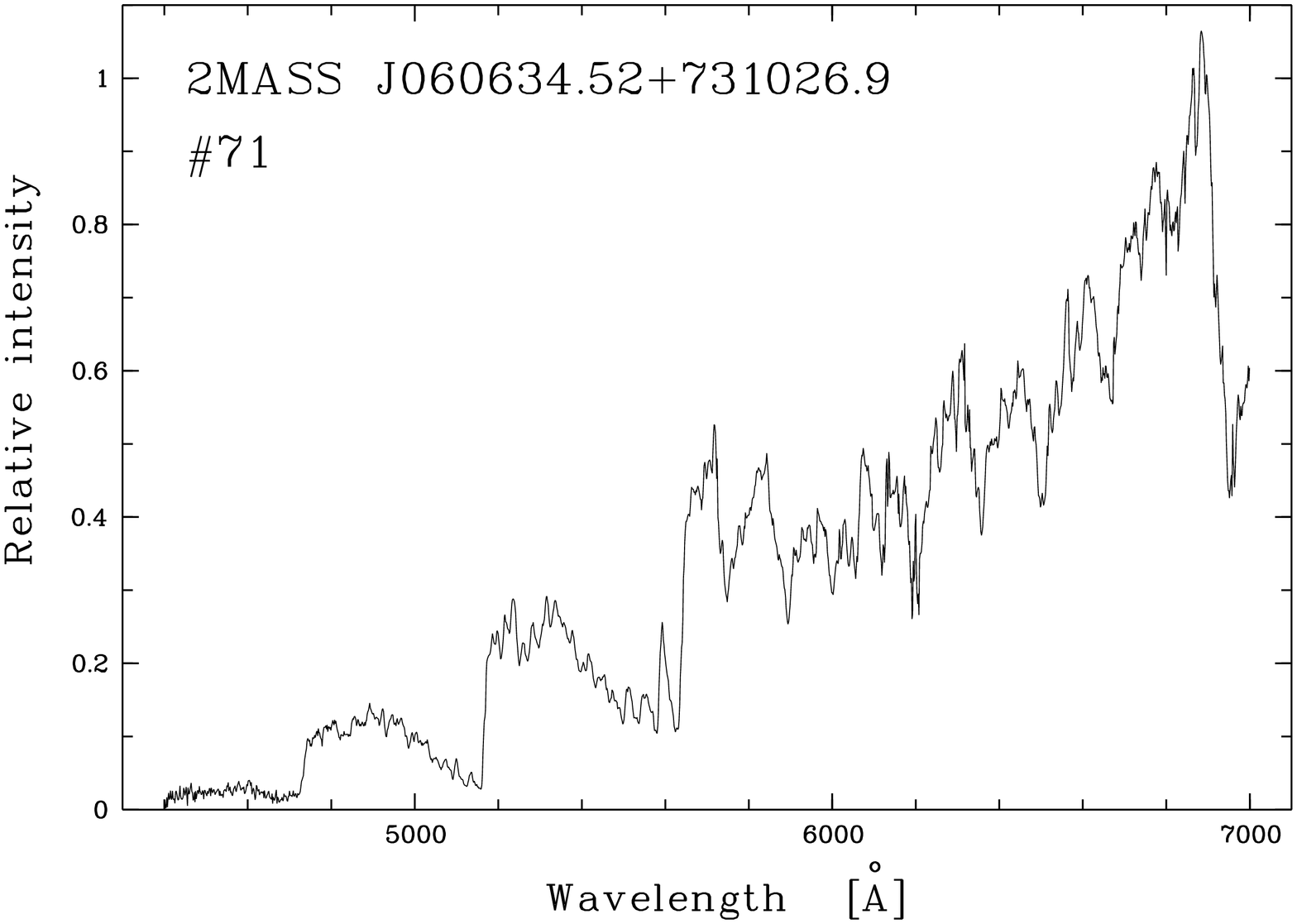}}}
\resizebox{8.5cm}{!}{\rotatebox{-90}{\includegraphics{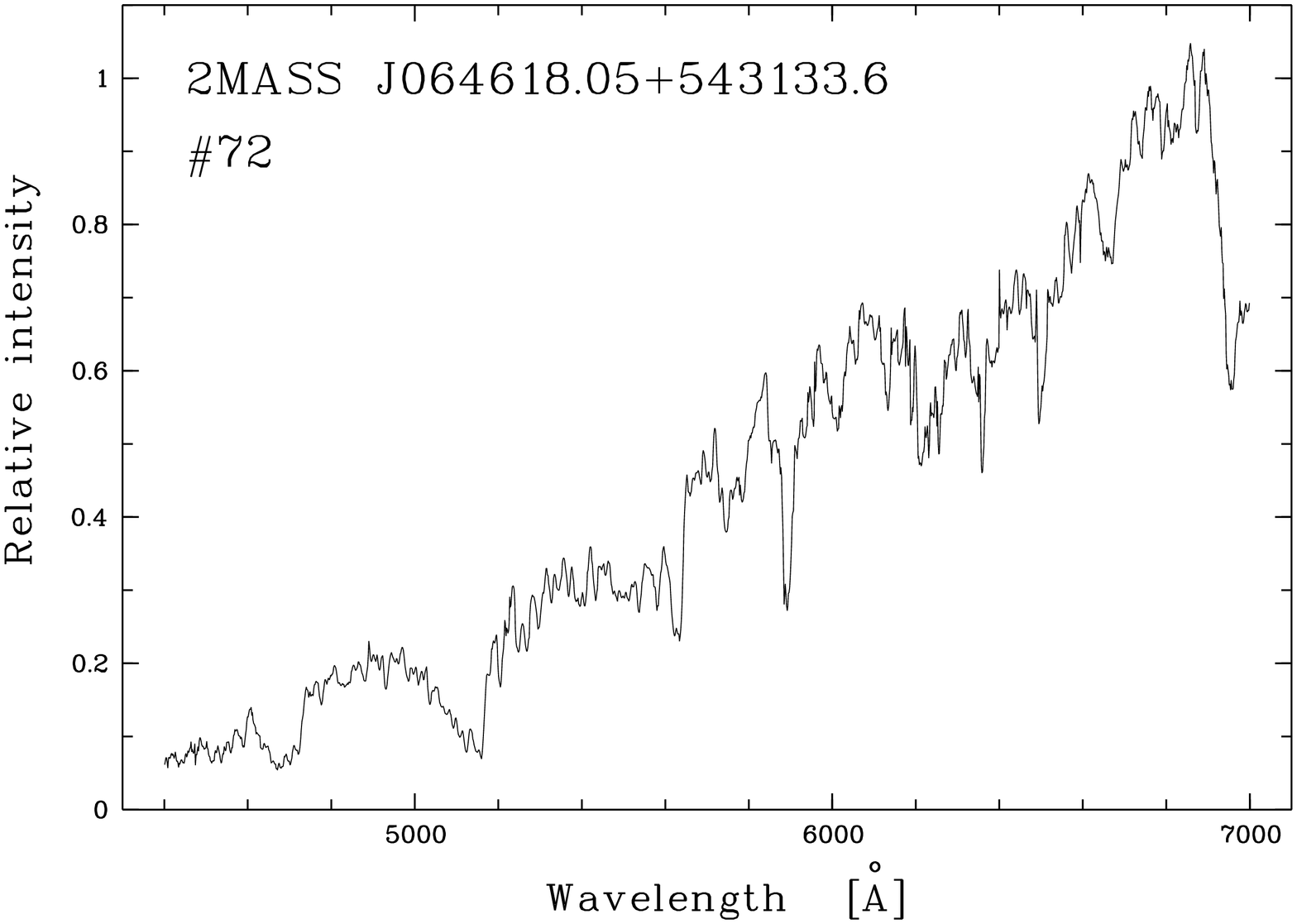}}}
\resizebox{8.5cm}{!}{\rotatebox{-90}{\includegraphics{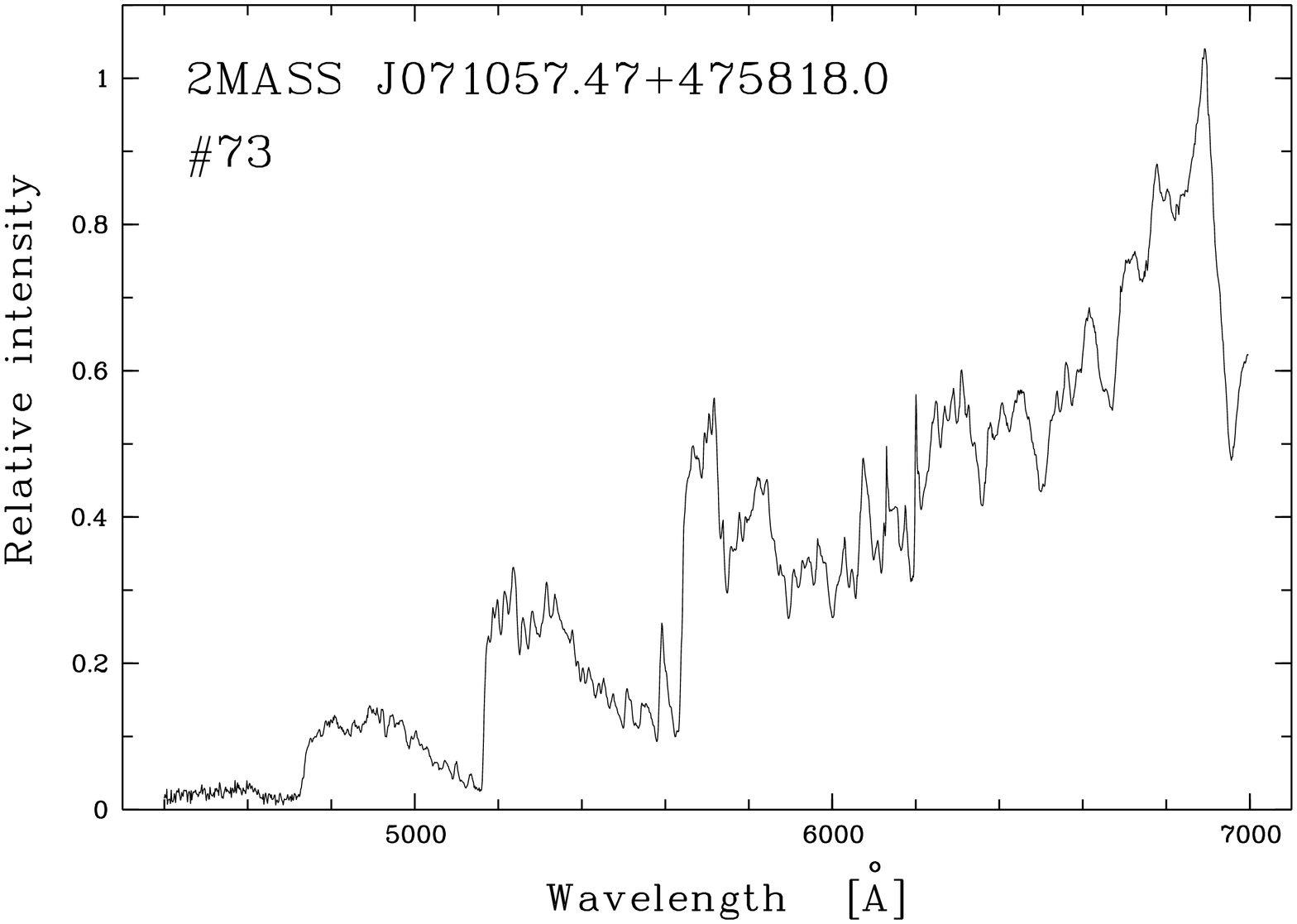}}}
\end{figure*}

\begin{figure*}
\caption[]{Stars observed at Byurakan (continued)}
\resizebox{8.5cm}{!}{\rotatebox{-90}{\includegraphics{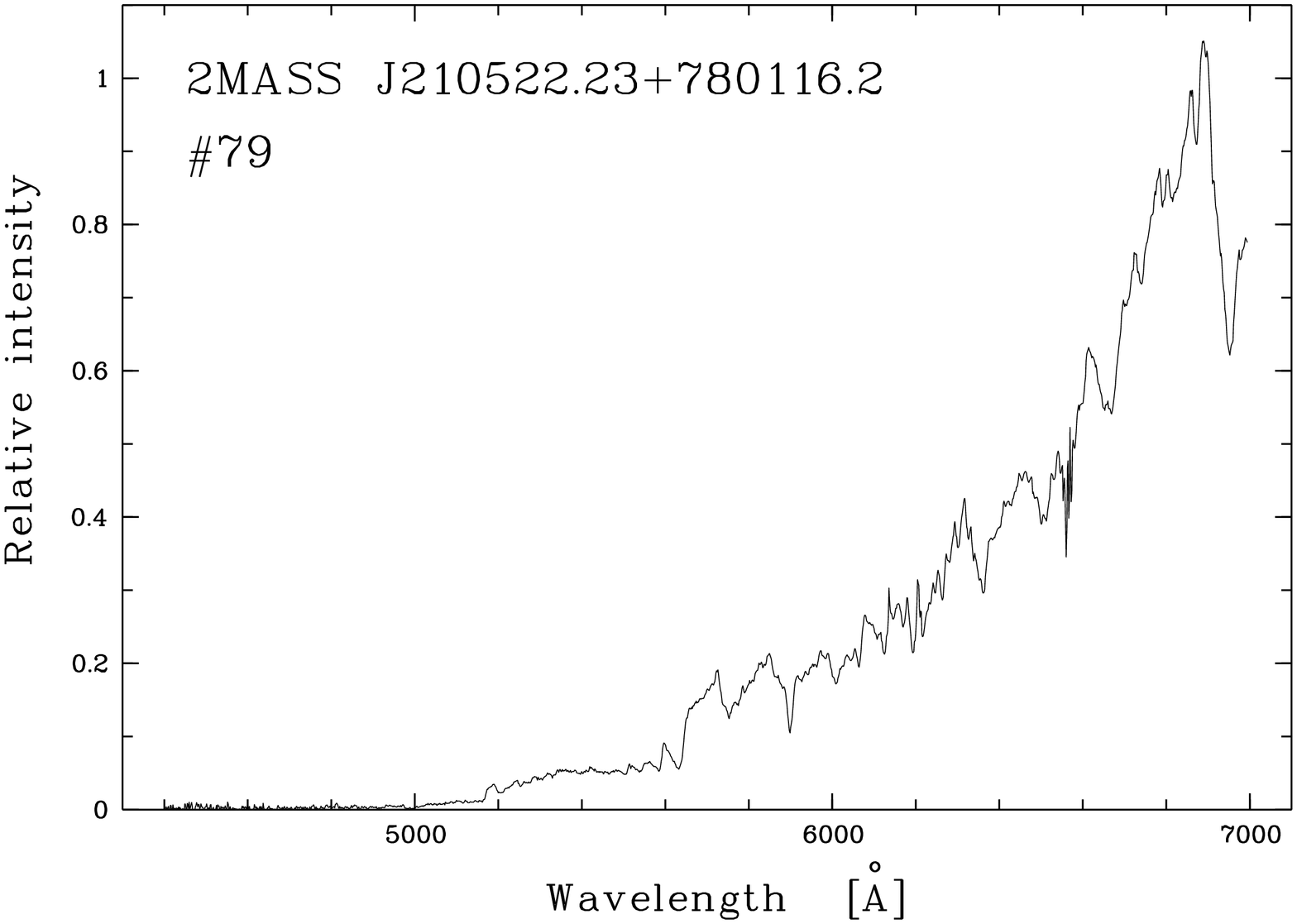}}}
\resizebox{8.5cm}{!}{\rotatebox{-90}{\includegraphics{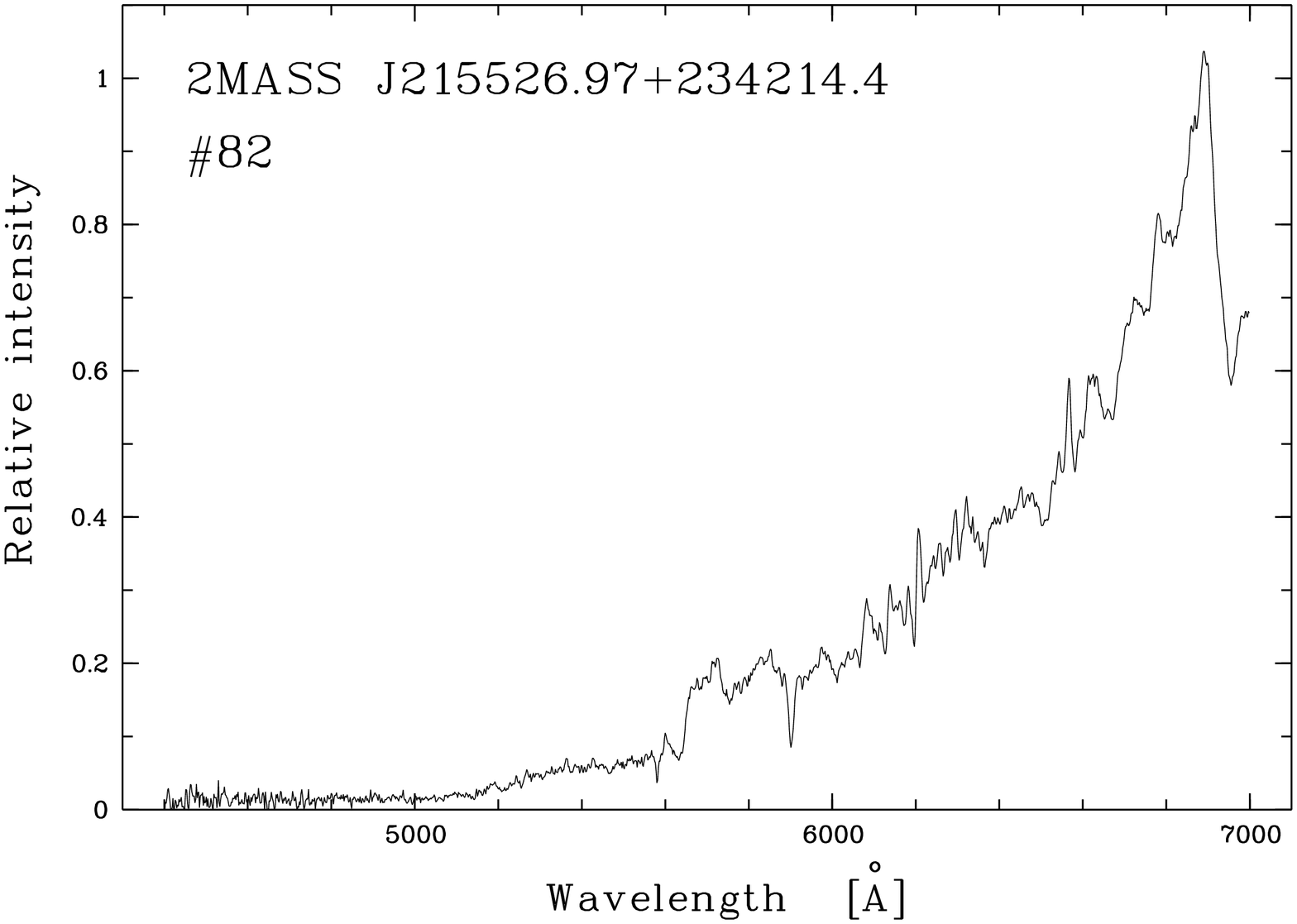}}}
\end{figure*}
\begin{figure*}
\caption[]{ Stars observed at OHP in September 2005}
\resizebox{8.5cm}{!}{\rotatebox{-90}{\includegraphics{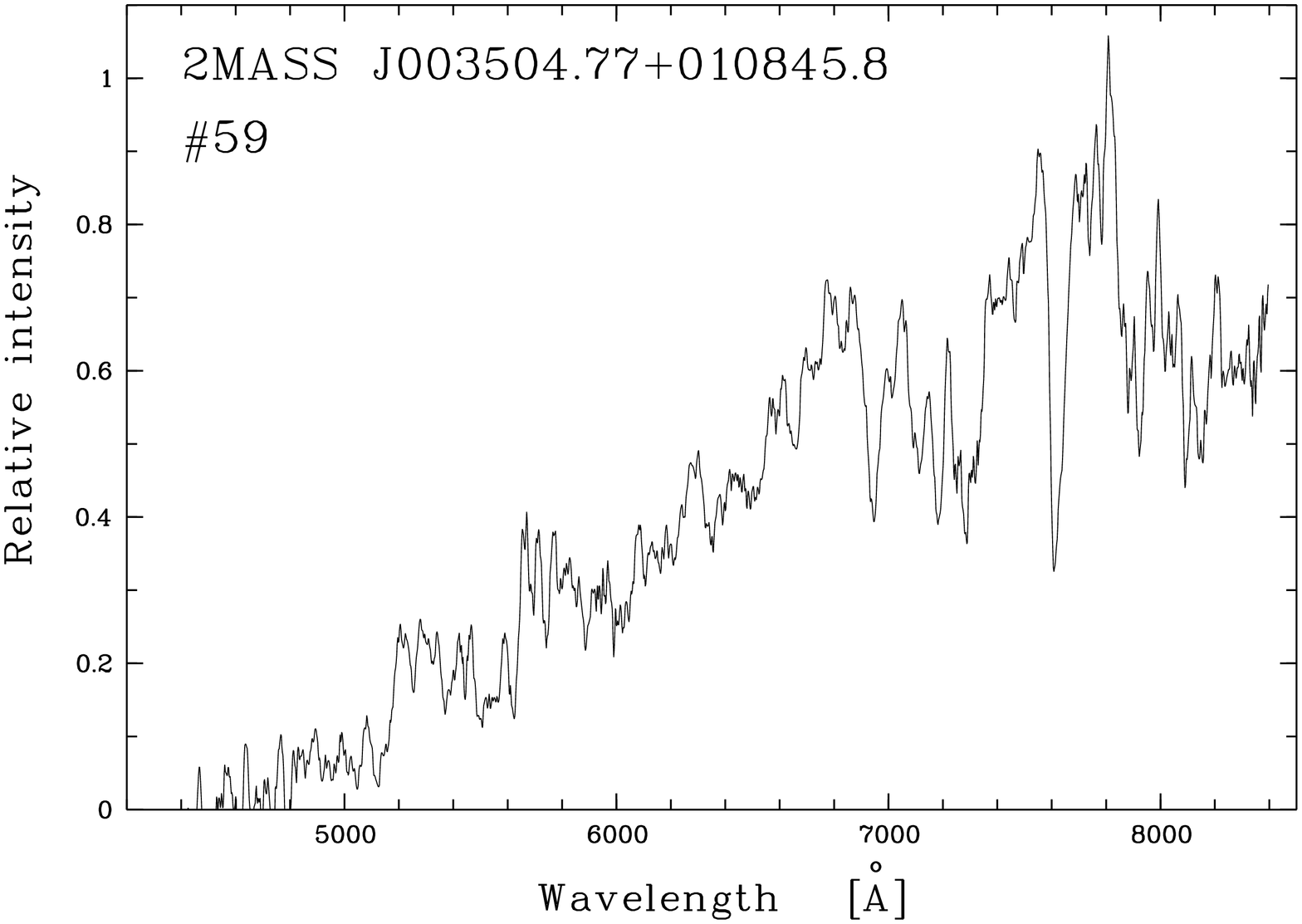}}}
\resizebox{8.5cm}{!}{\rotatebox{-90}{\includegraphics{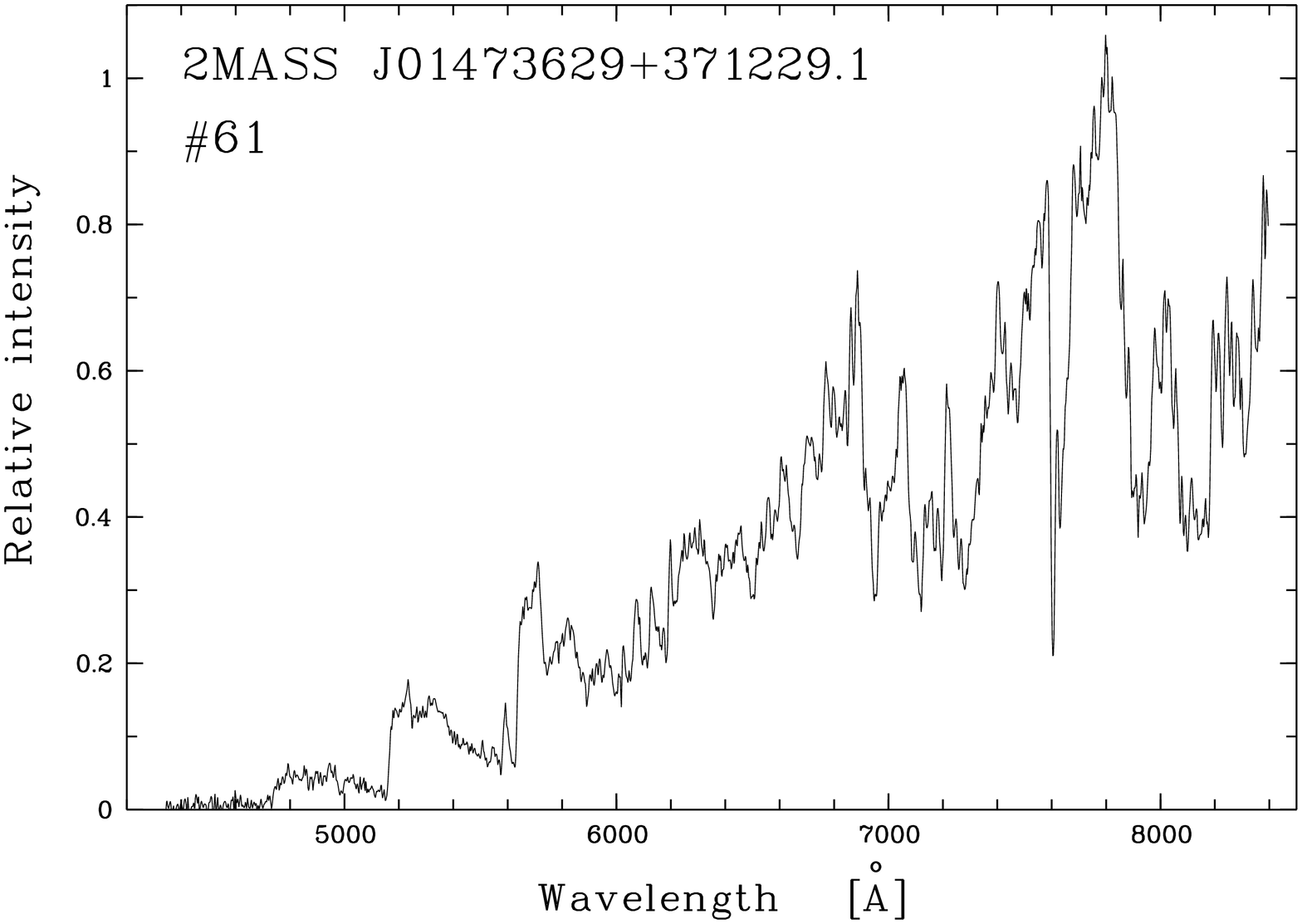}}}
\resizebox{8.5cm}{!}{\rotatebox{-90}{\includegraphics{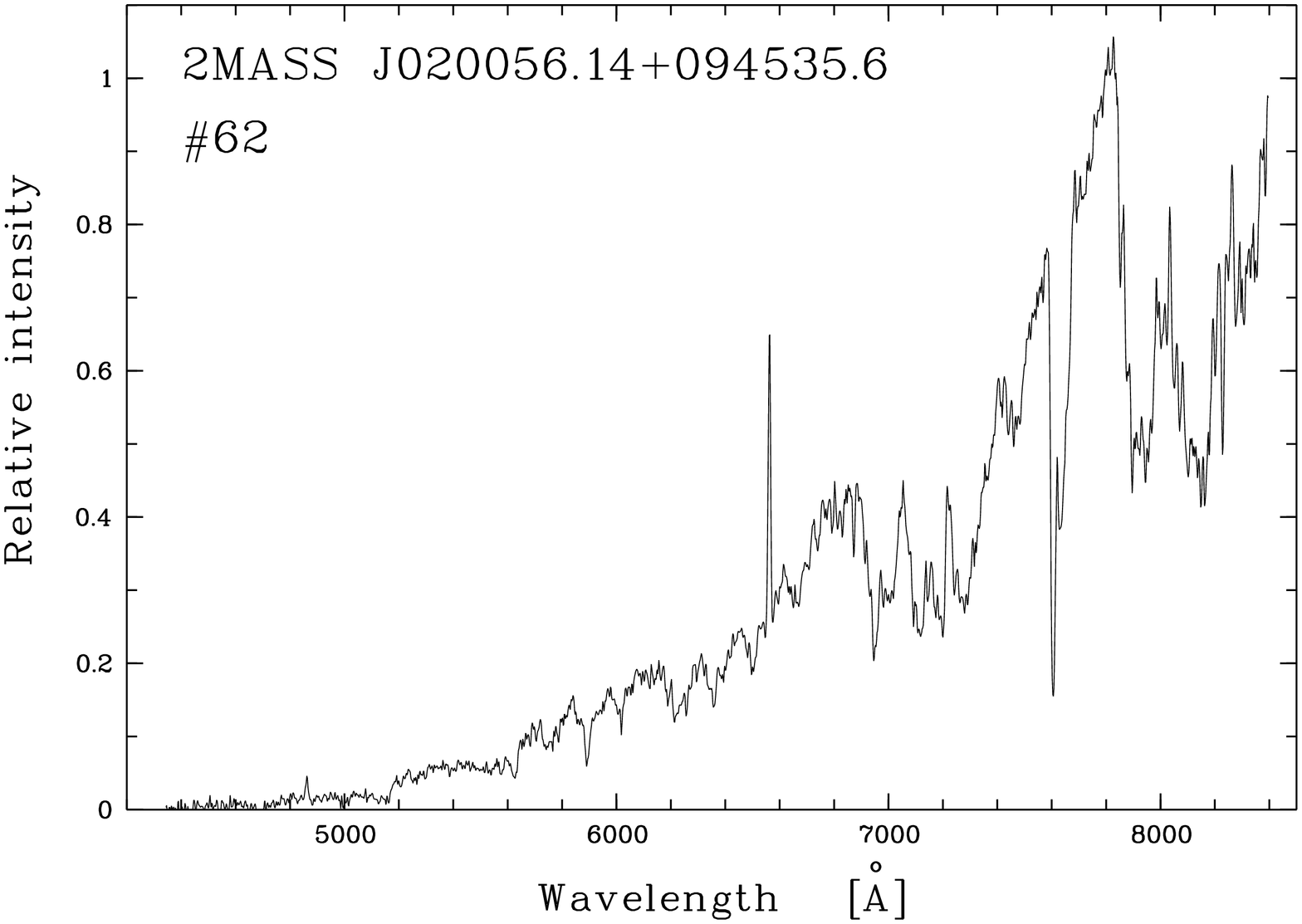}}}
\resizebox{8.5cm}{!}{\rotatebox{-90}{\includegraphics{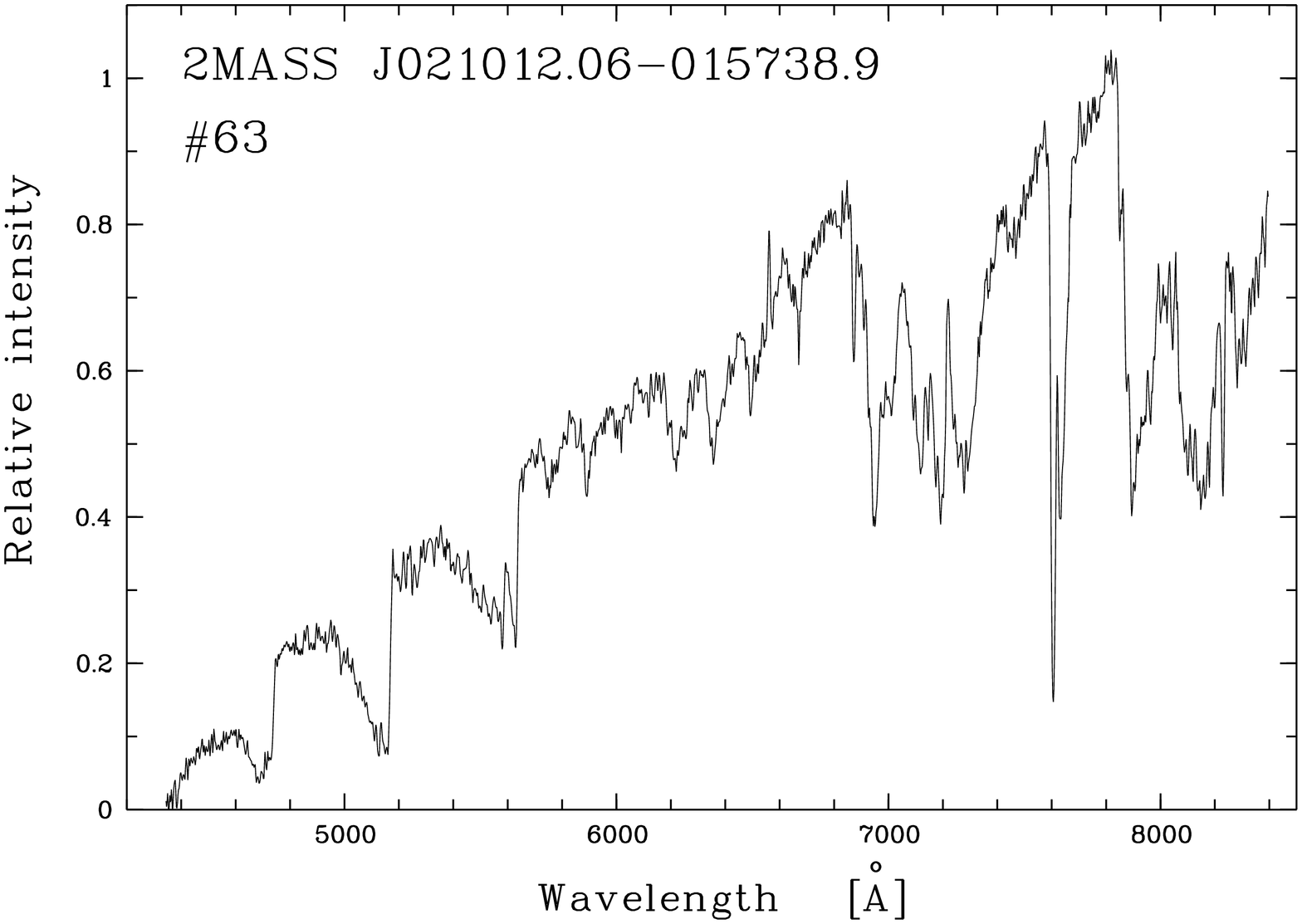}}}
\resizebox{8.5cm}{!}{\rotatebox{-90}{\includegraphics{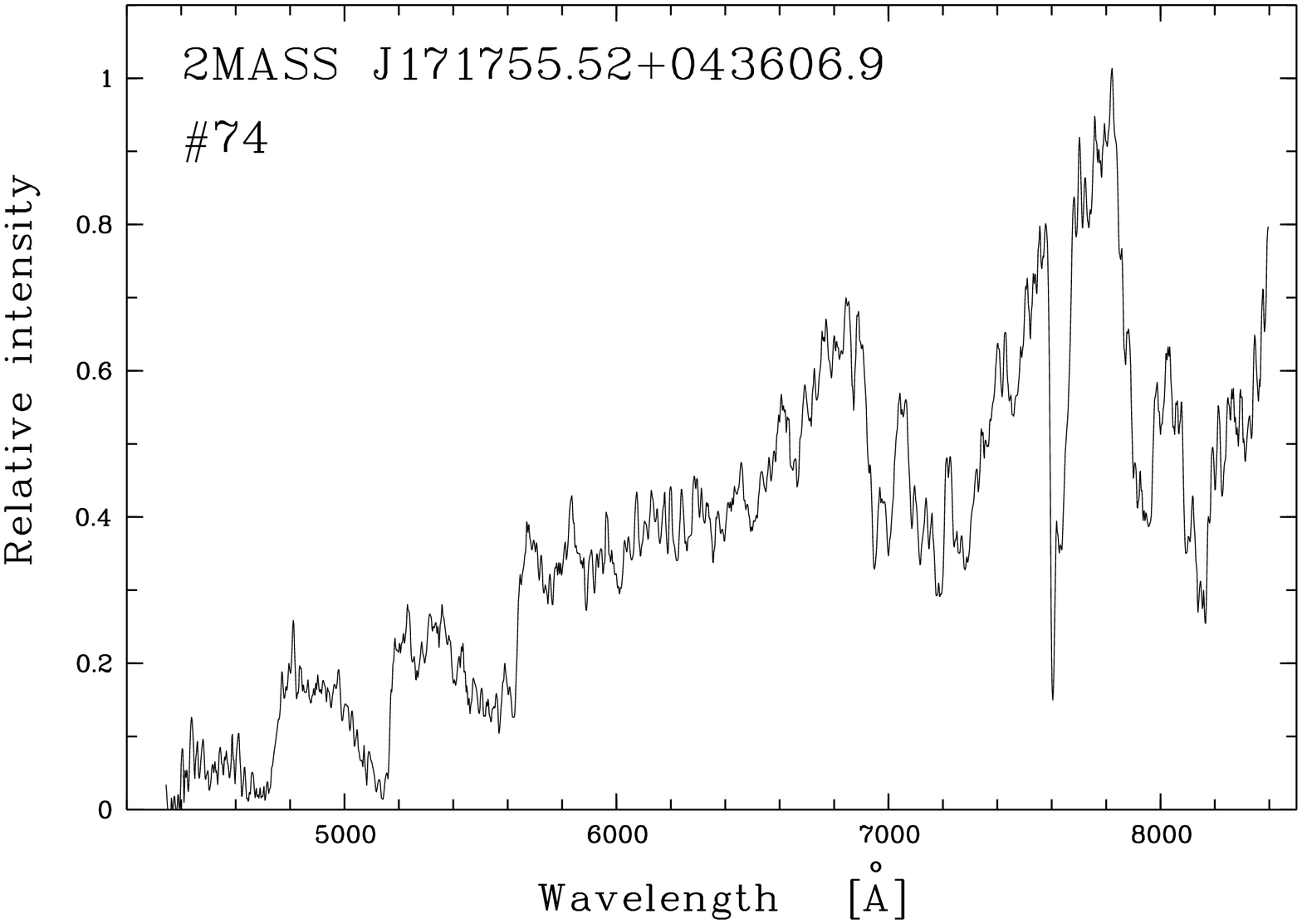}}}
\resizebox{8.5cm}{!}{\rotatebox{-90}{\includegraphics{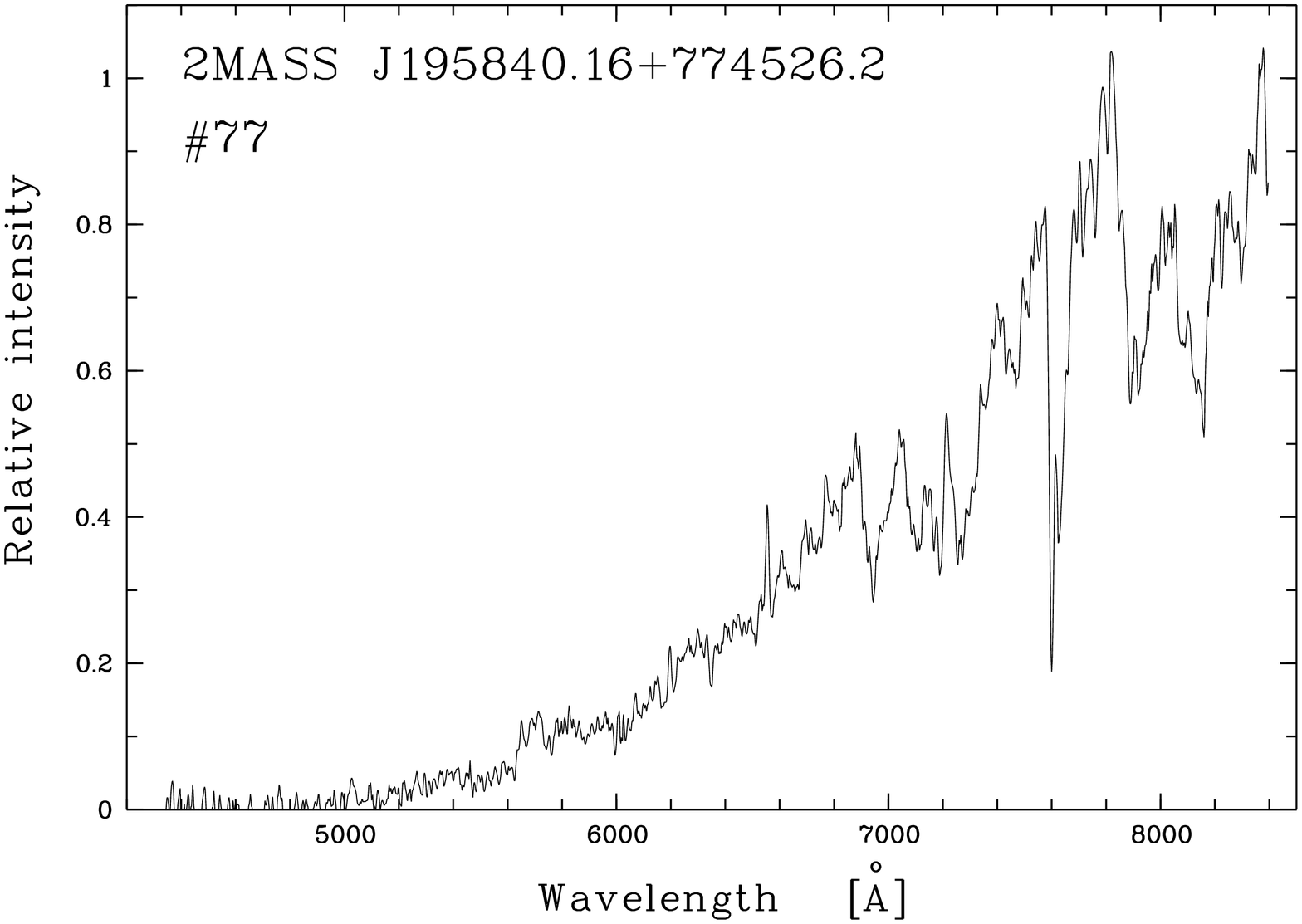}}}
\resizebox{8.5cm}{!}{\rotatebox{-90}{\includegraphics{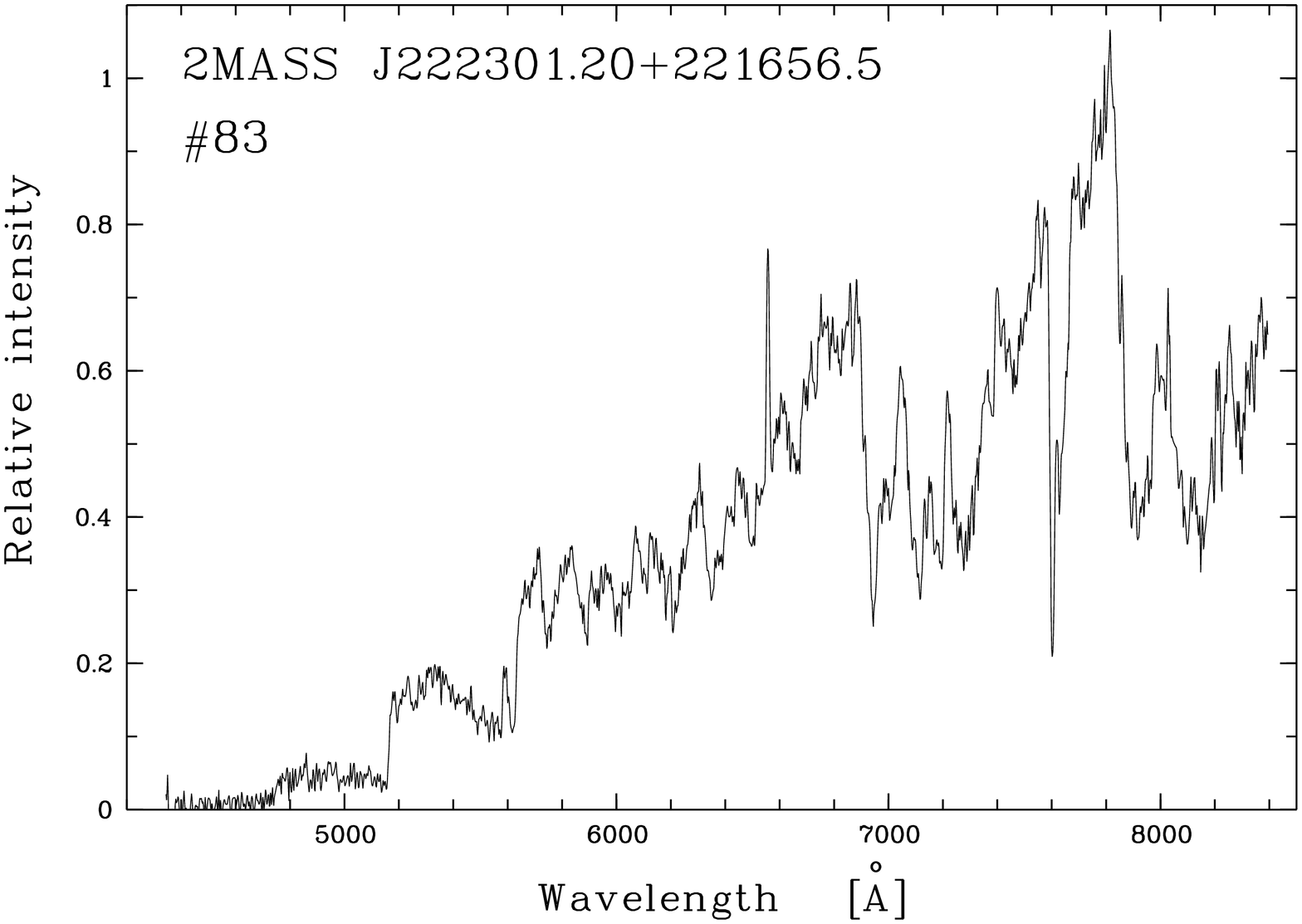}}}
\end{figure*}

\begin{figure*}
\caption[]{Stars observed at OHP in September 2006}
\resizebox{8.5cm}{!}{\rotatebox{-90}{\includegraphics{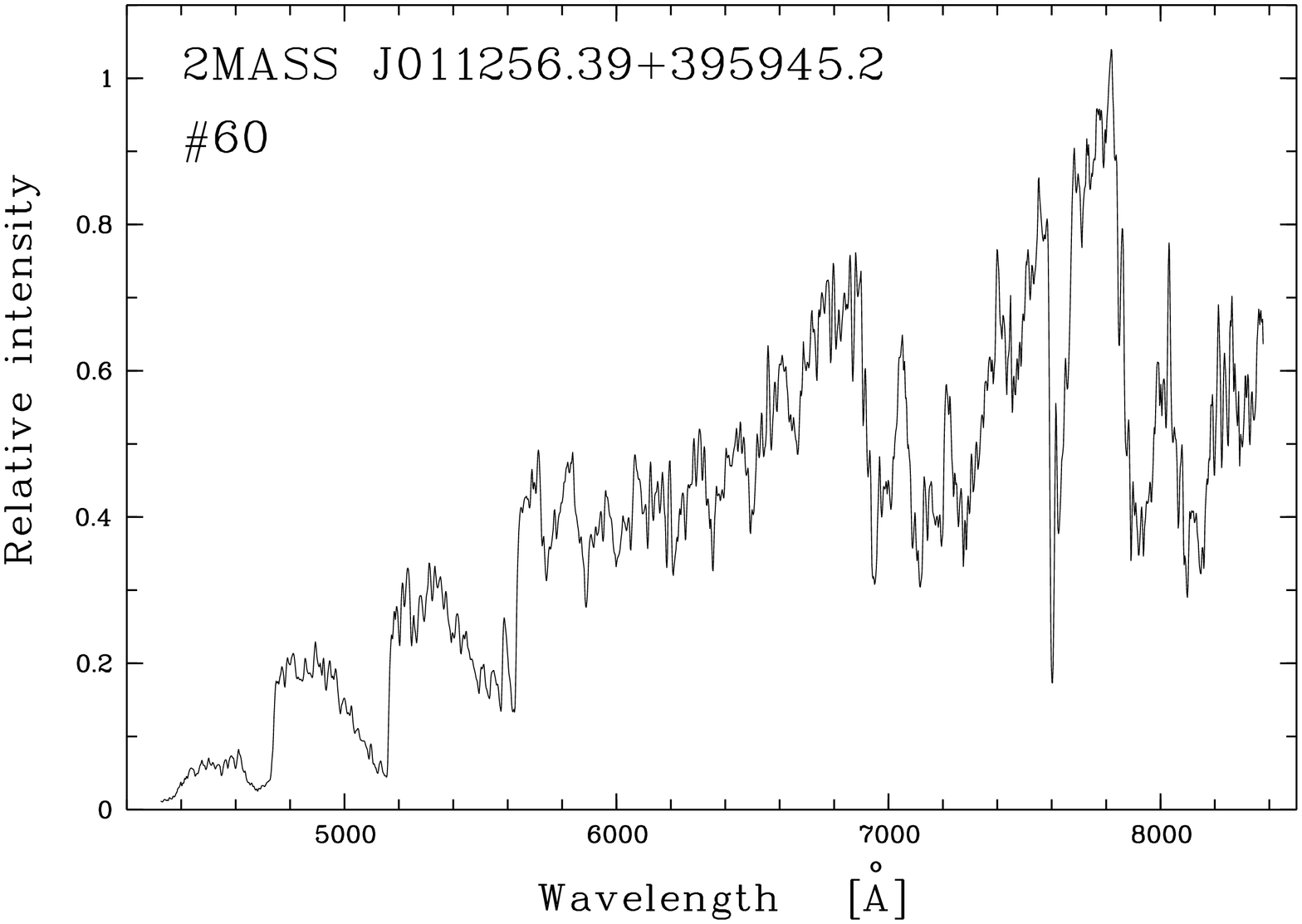}}}
\resizebox{8.5cm}{!}{\rotatebox{-90}{\includegraphics{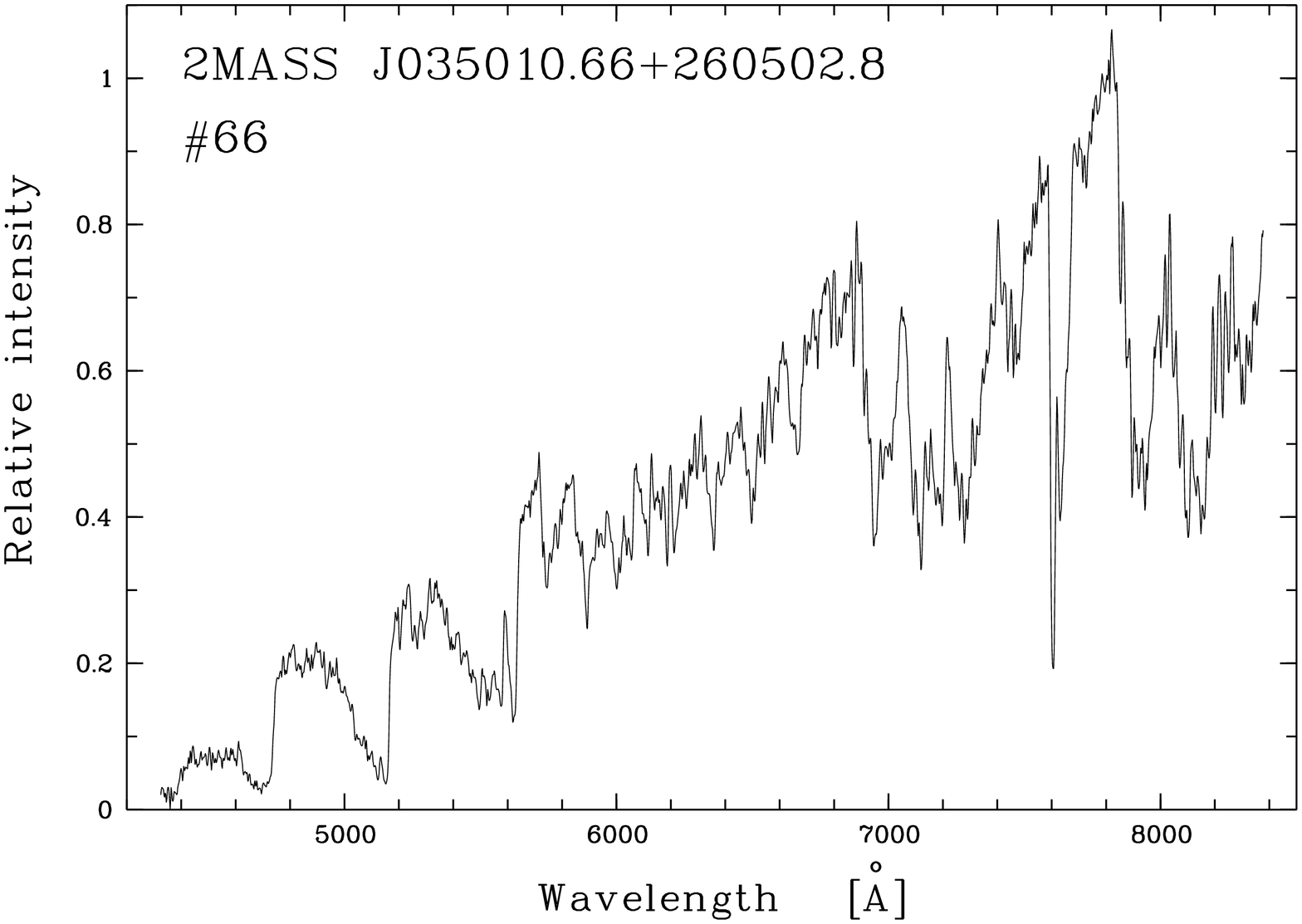}}}
\resizebox{8.5cm}{!}{\rotatebox{-90}{\includegraphics{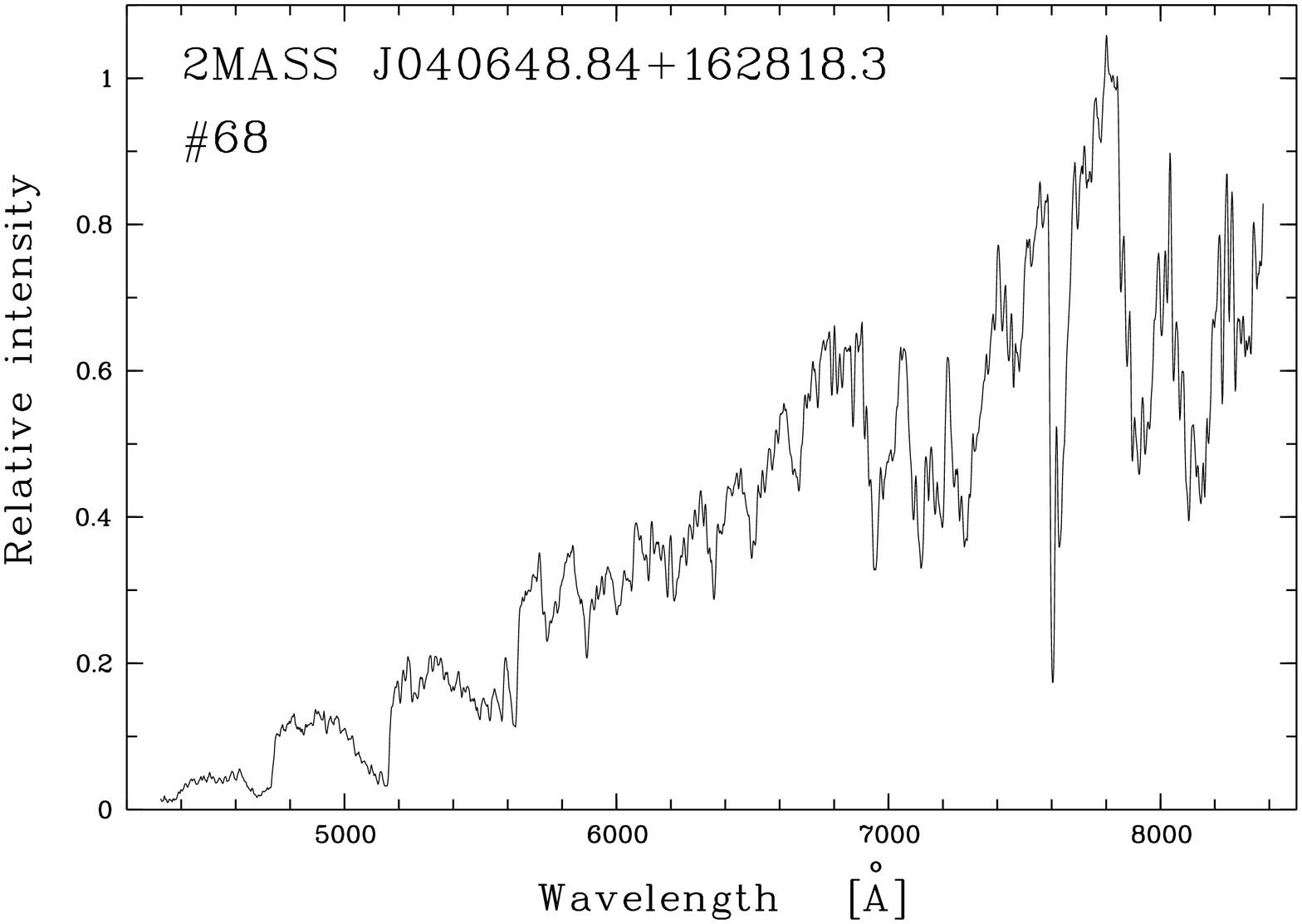}}}
\resizebox{8.5cm}{!}{\rotatebox{-90}{\includegraphics{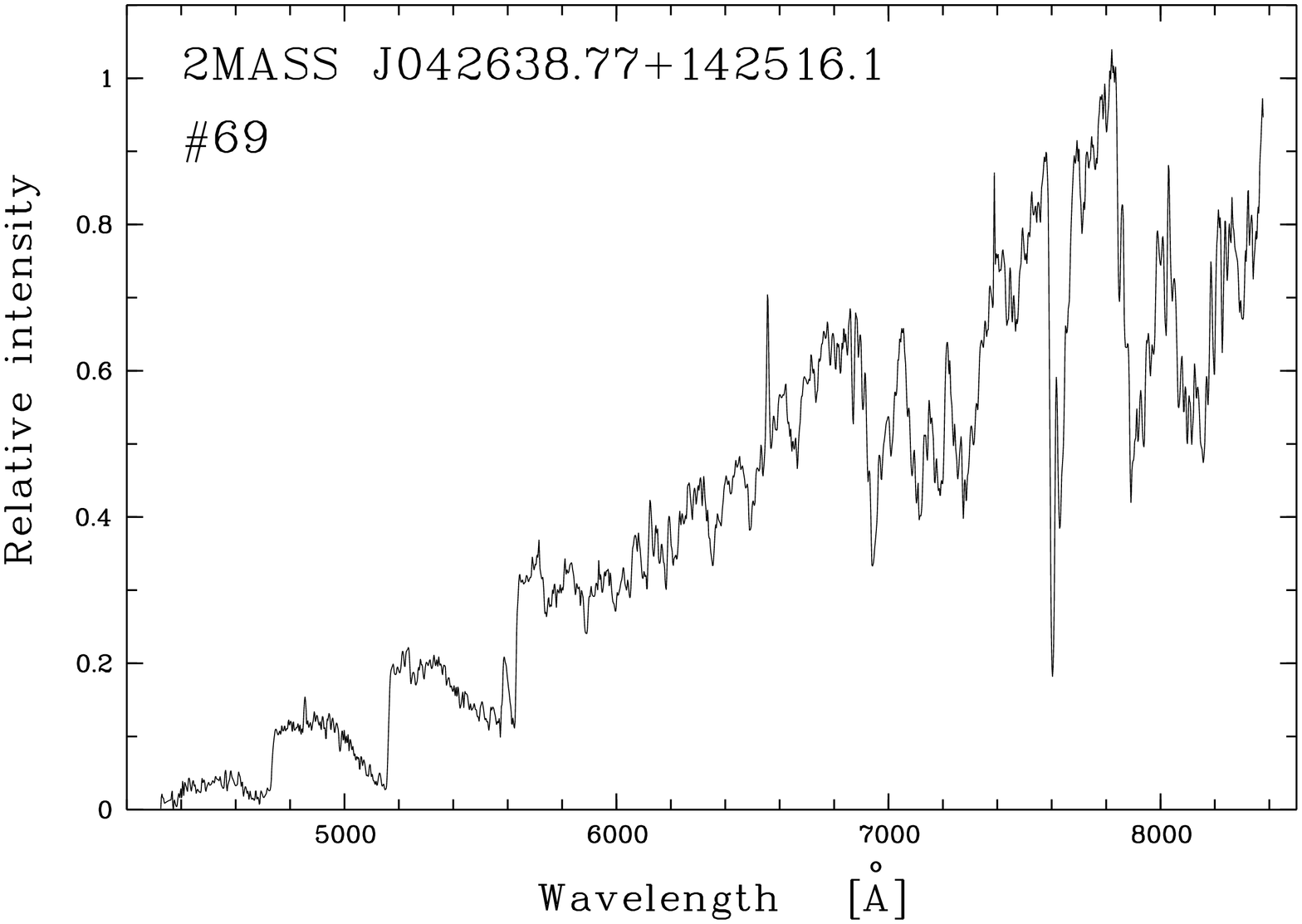}}}
\resizebox{8.5cm}{!}{\rotatebox{-90}{\includegraphics{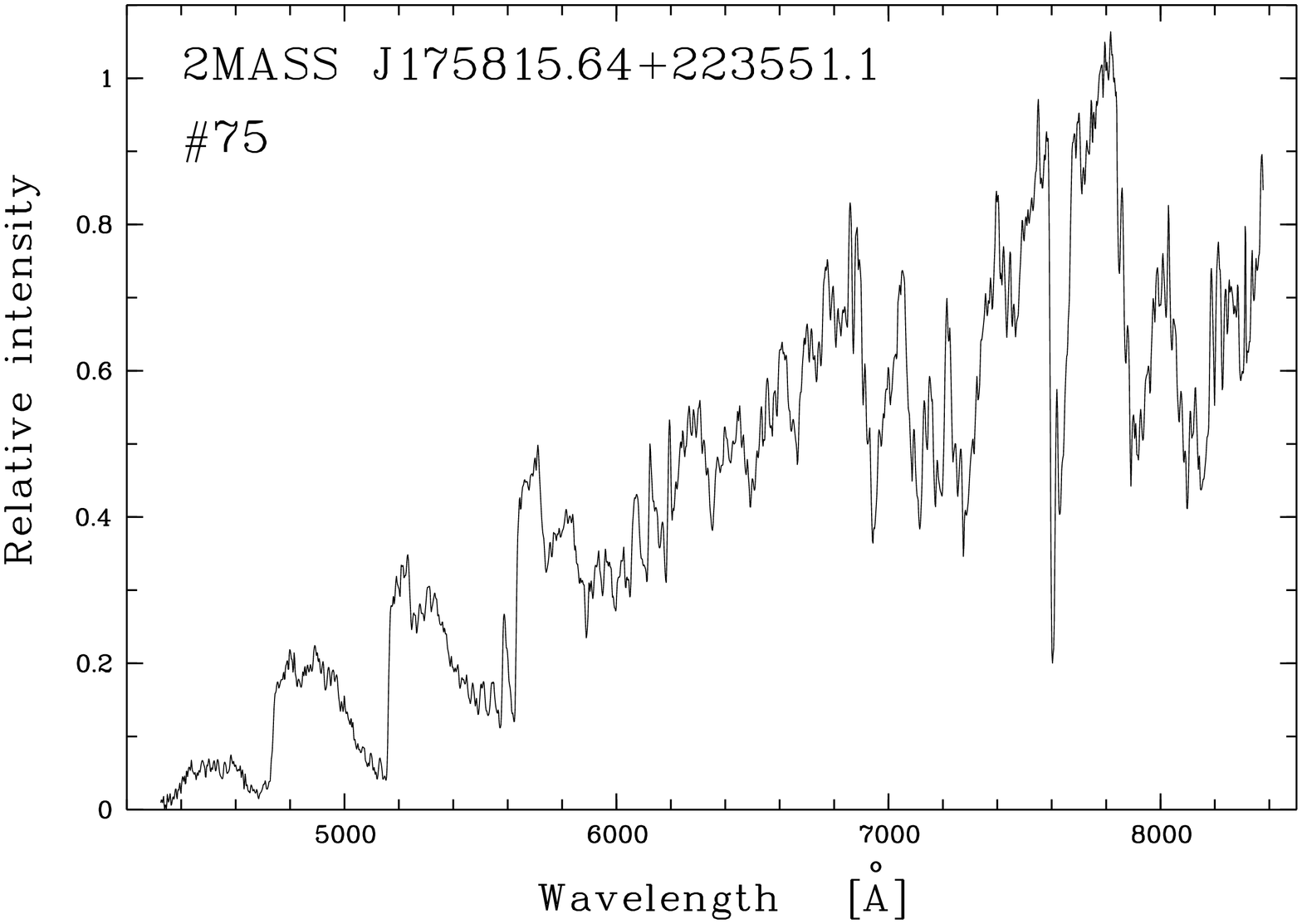}}}
\resizebox{8.5cm}{!}{\rotatebox{-90}{\includegraphics{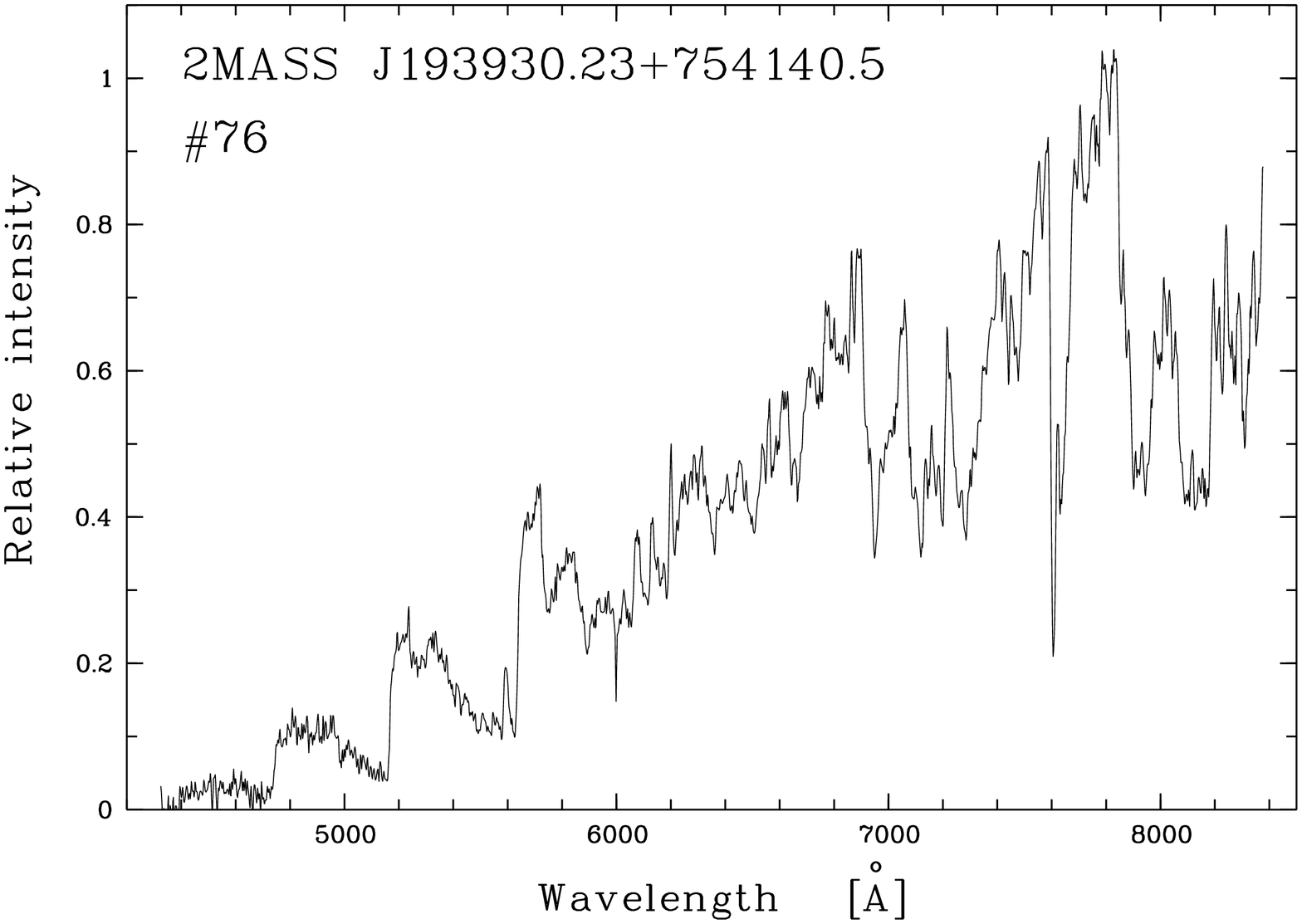}}}
\resizebox{8.5cm}{!}{\rotatebox{-90}{\includegraphics{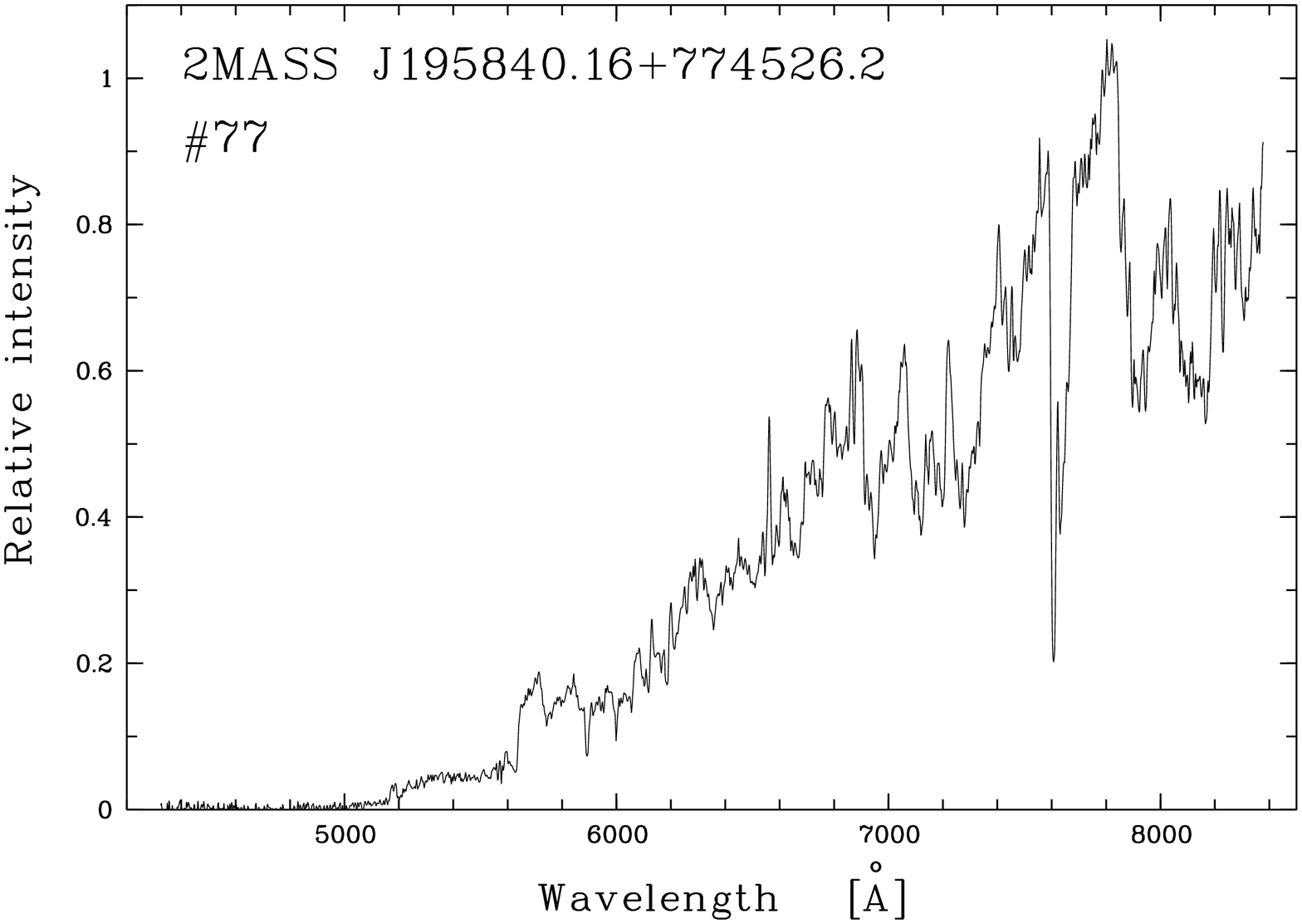}}}
\resizebox{8.5cm}{!}{\rotatebox{-90}{\includegraphics{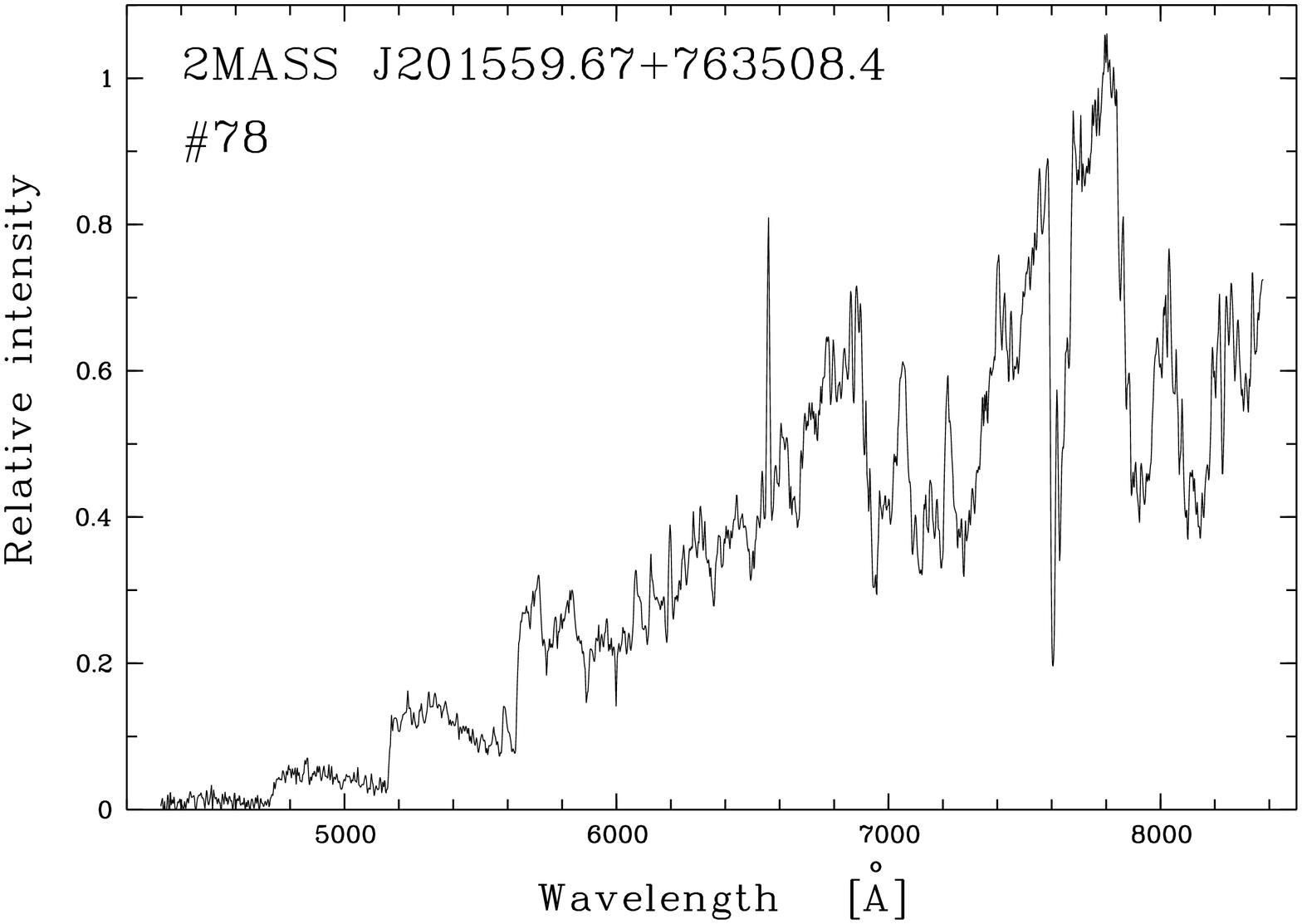}}}
\end{figure*}

\begin{figure*}
\caption[]{Stars observed at OHP in September 2006 (continued)}
\resizebox{8.5cm}{!}{\rotatebox{-90}{\includegraphics{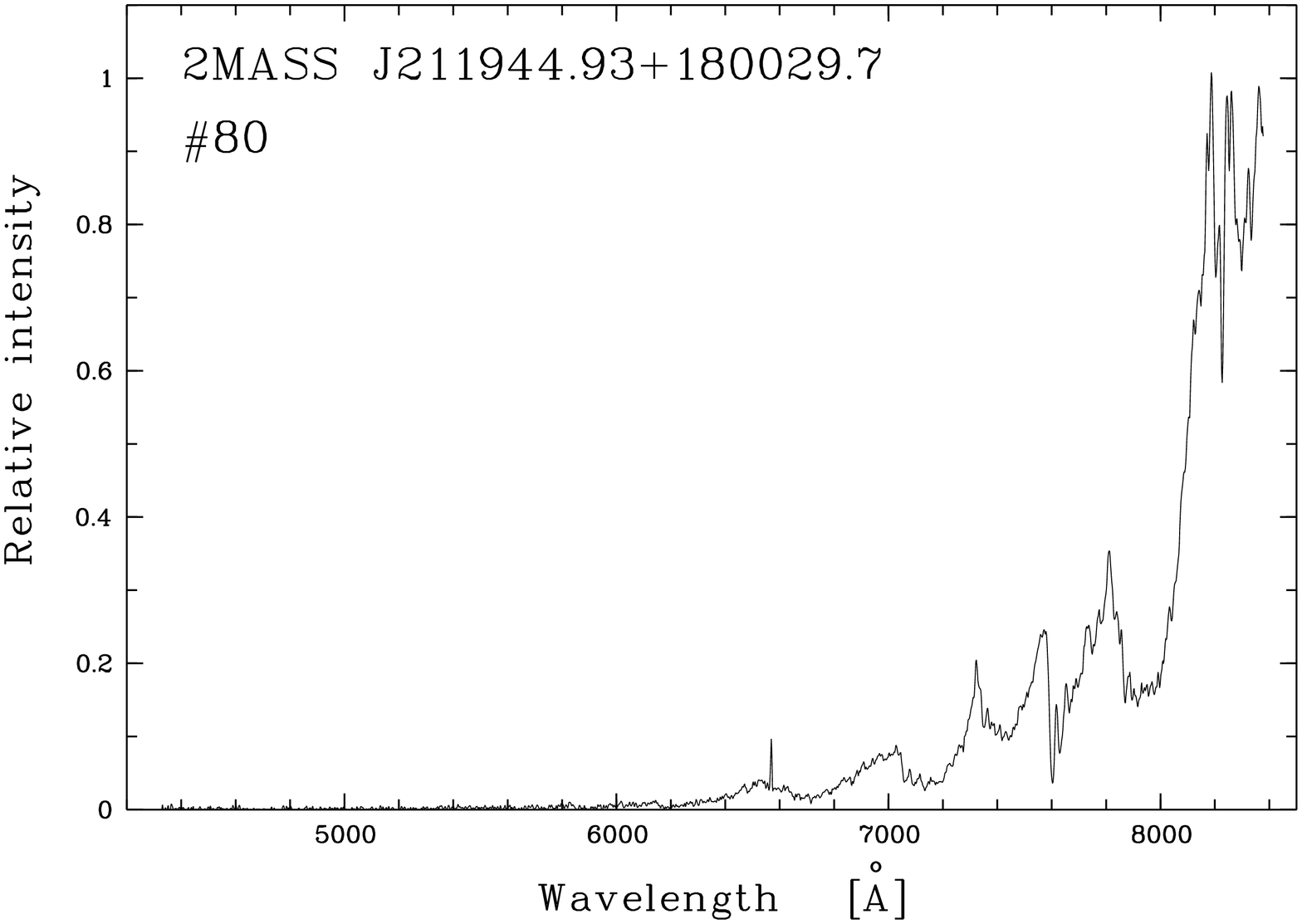}}}
\resizebox{8.5cm}{!}{\rotatebox{-90}{\includegraphics{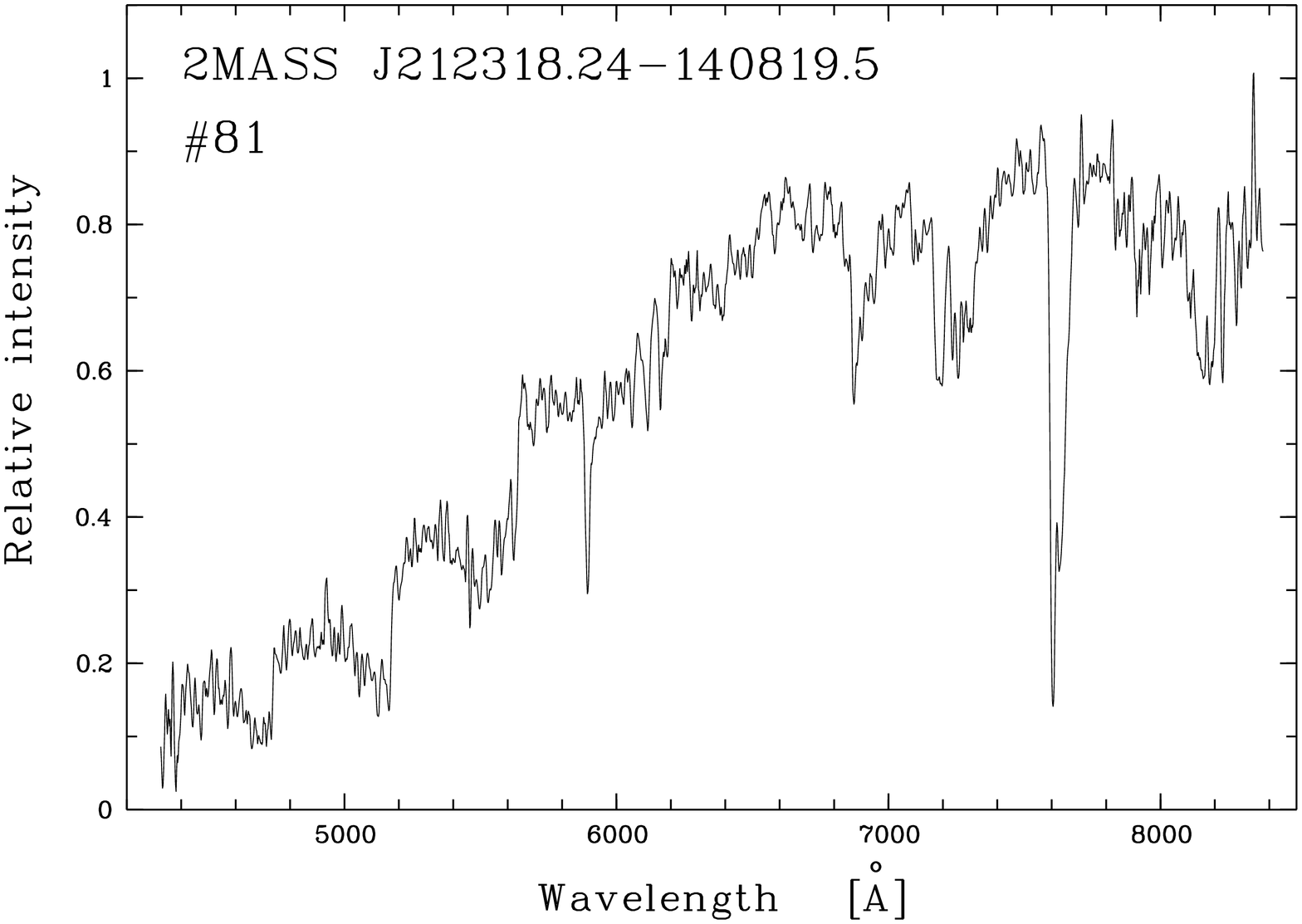}}}
\end{figure*}

\begin{figure*}
\caption[]{Light curves from NSVS data}
\resizebox{8.5cm}{!}{\rotatebox{-90}{\includegraphics{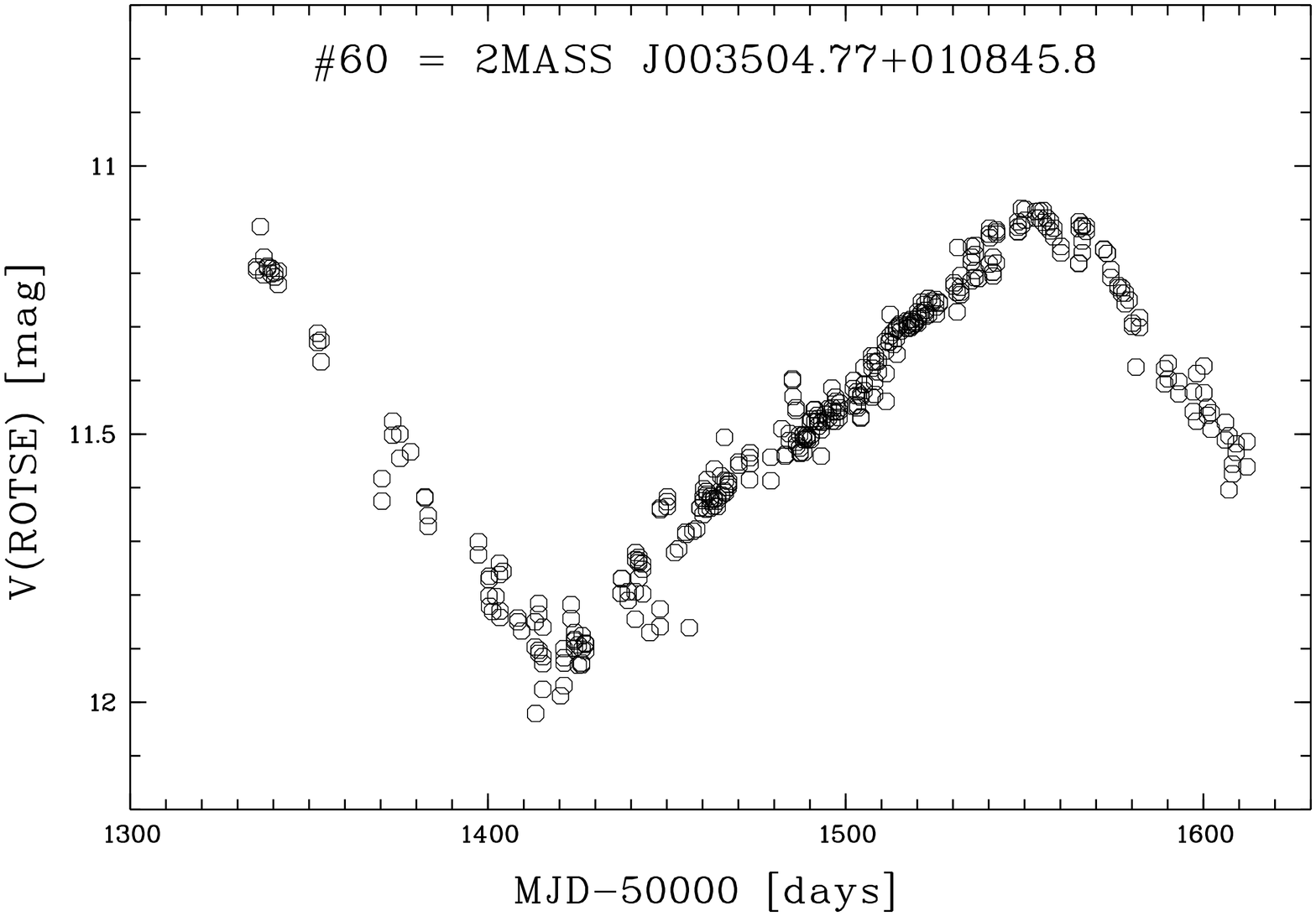}}}
\resizebox{8.5cm}{!}{\rotatebox{-90}{\includegraphics{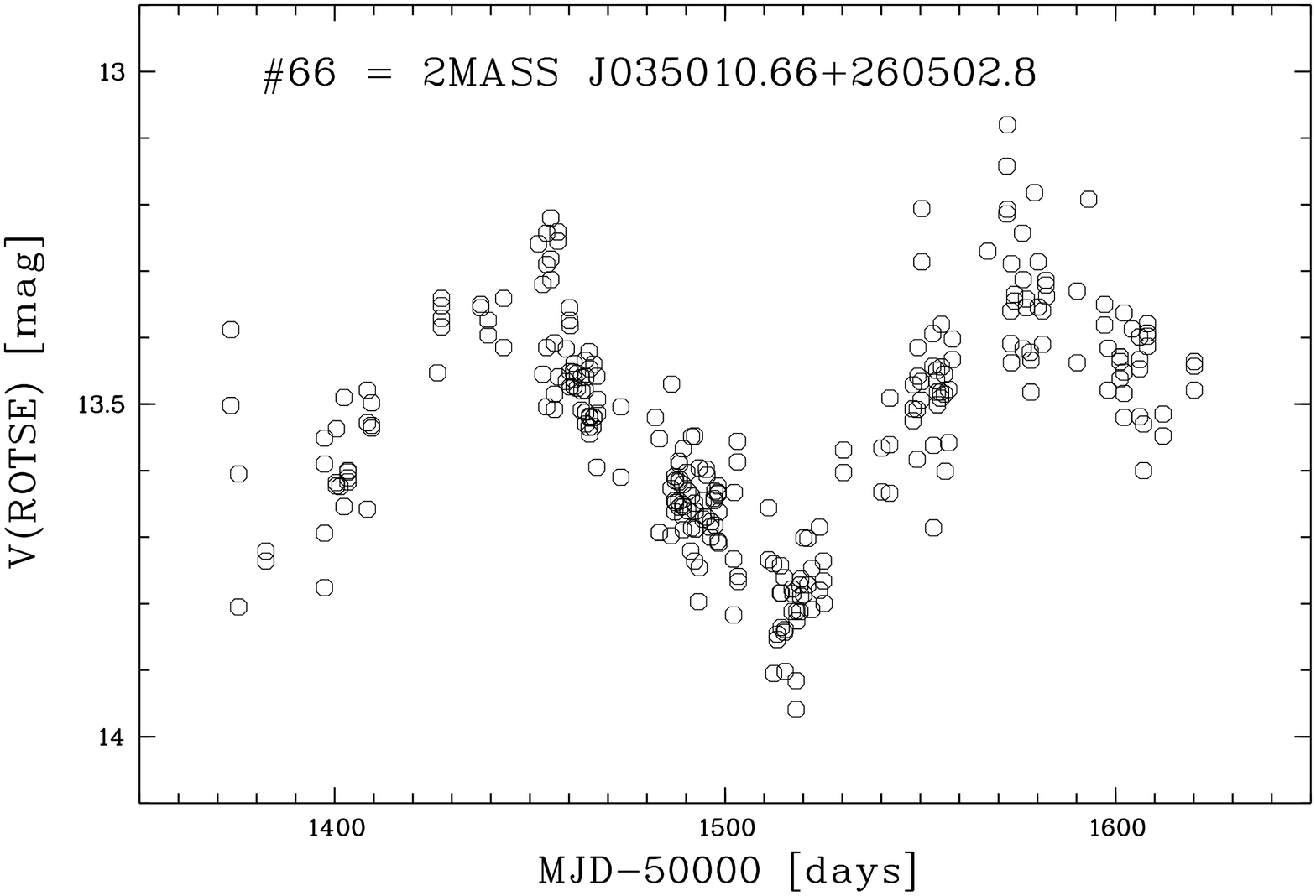}}}
\resizebox{8.5cm}{!}{\rotatebox{-90}{\includegraphics{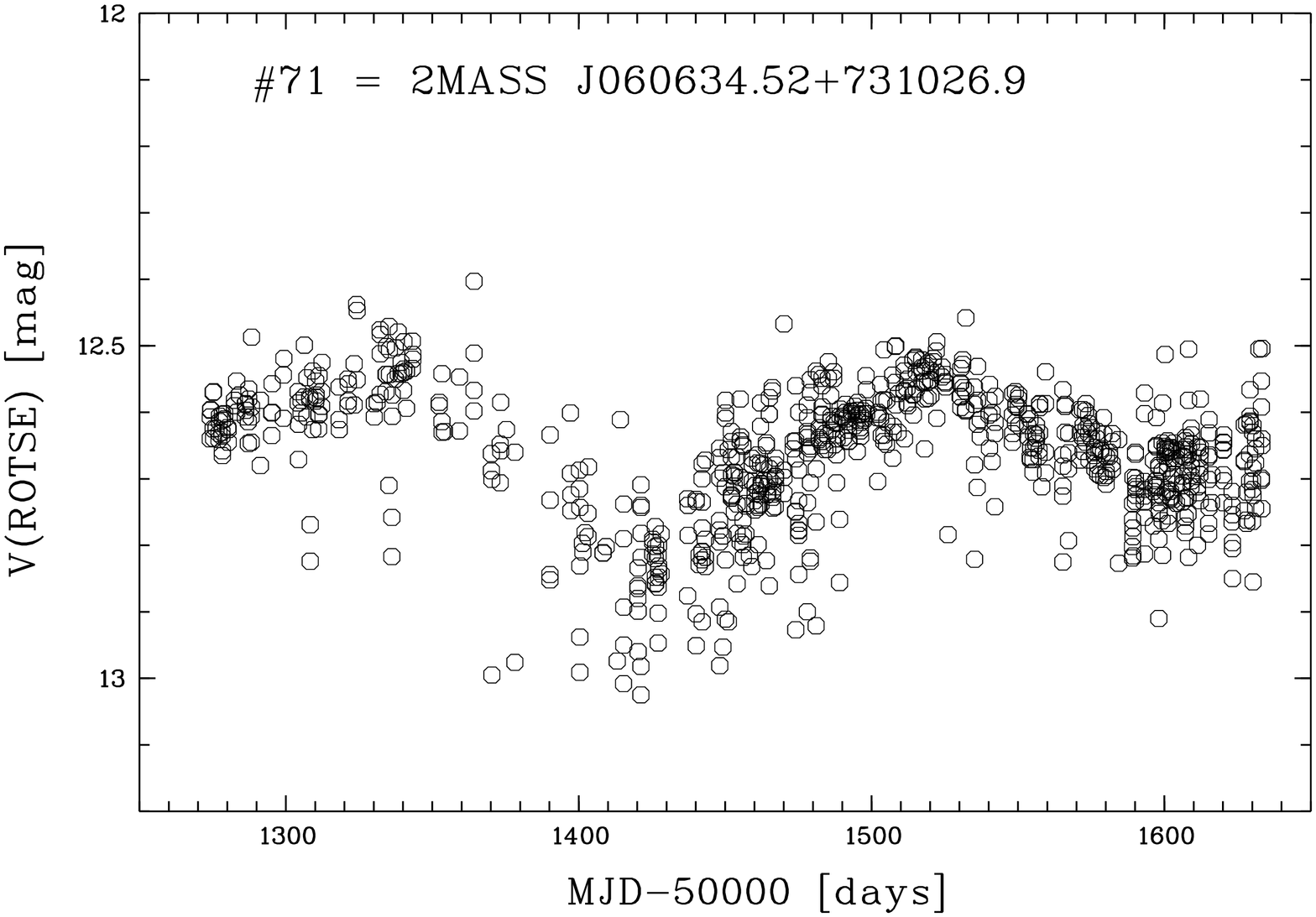}}}
\resizebox{8.5cm}{!}{\rotatebox{-90}{\includegraphics{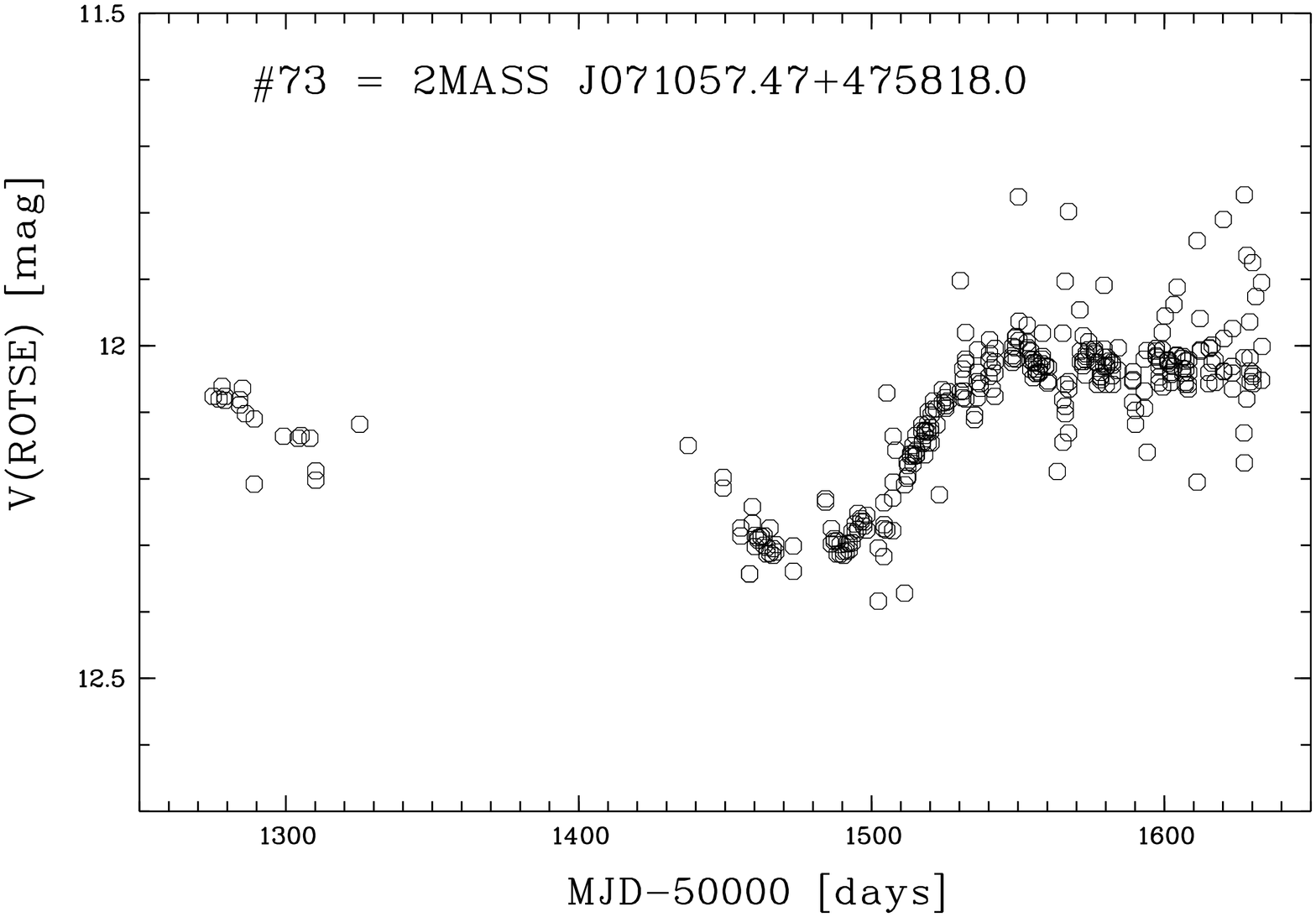}}}
\resizebox{8.5cm}{!}{\rotatebox{-90}{\includegraphics{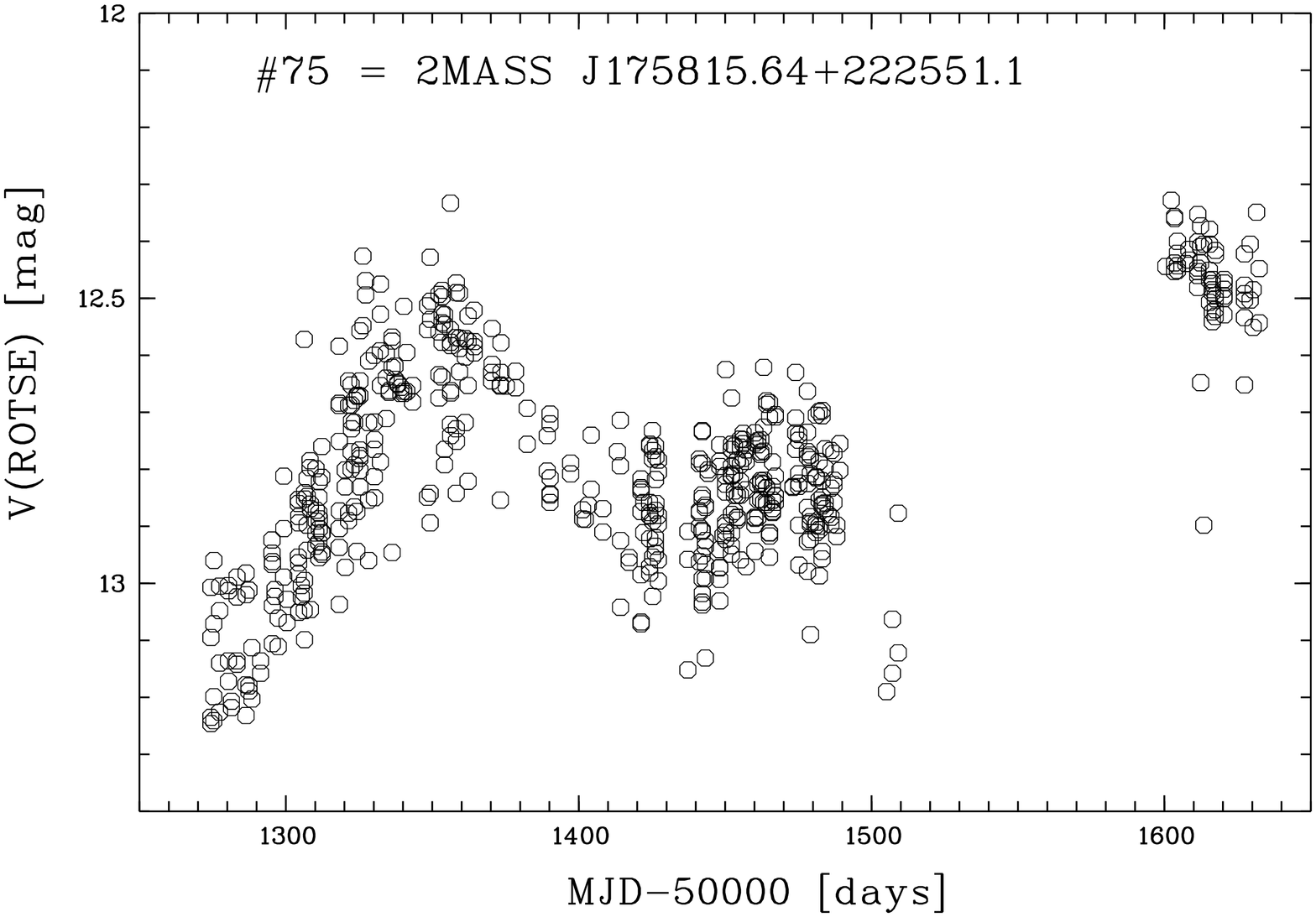}}}
\resizebox{8.5cm}{!}{\rotatebox{-90}{\includegraphics{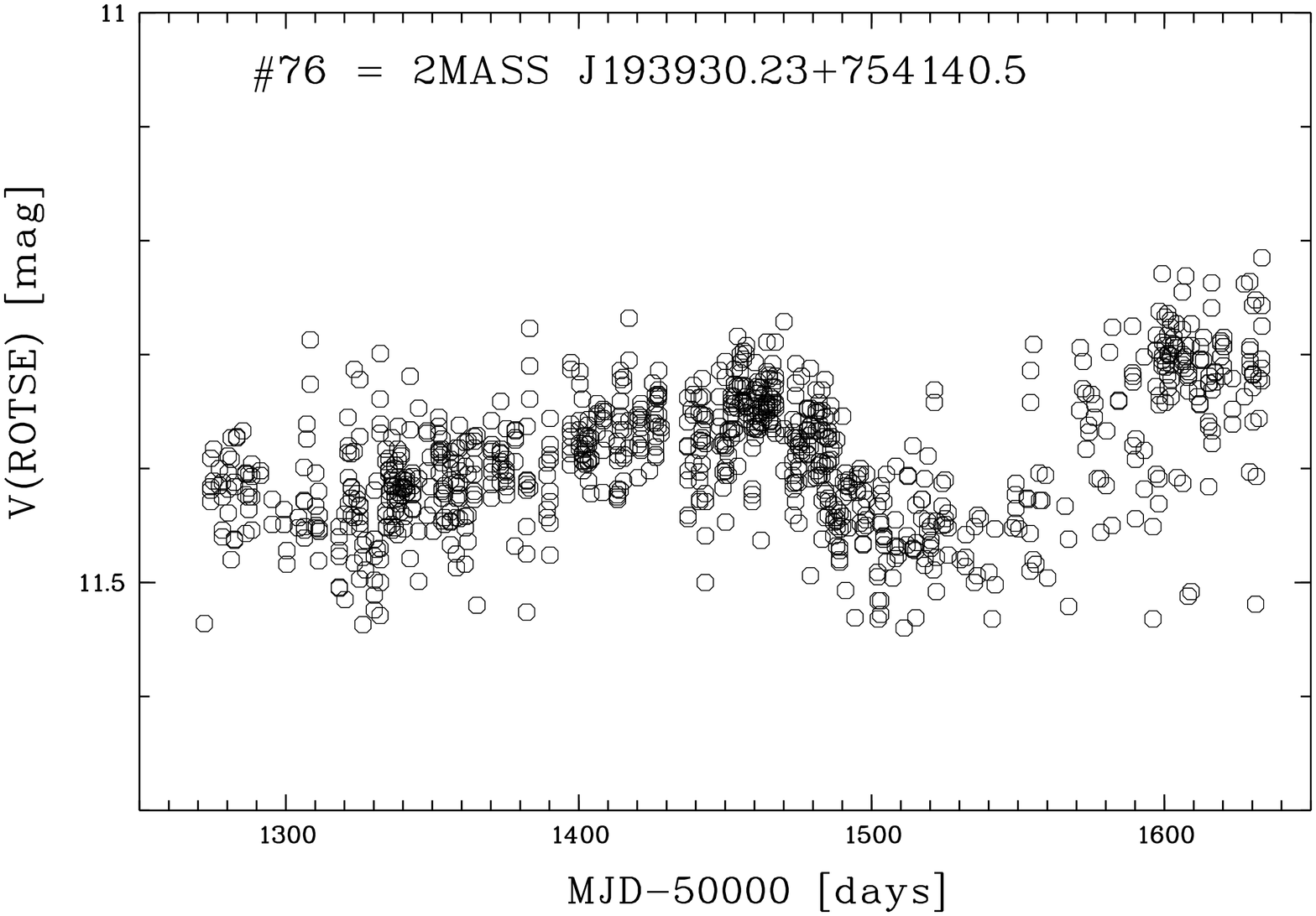}}}
\end{figure*}

\begin{figure*}
\caption[]{Light curves from NSVS data (continued)}
\resizebox{8.5cm}{!}{\rotatebox{-90}{\includegraphics{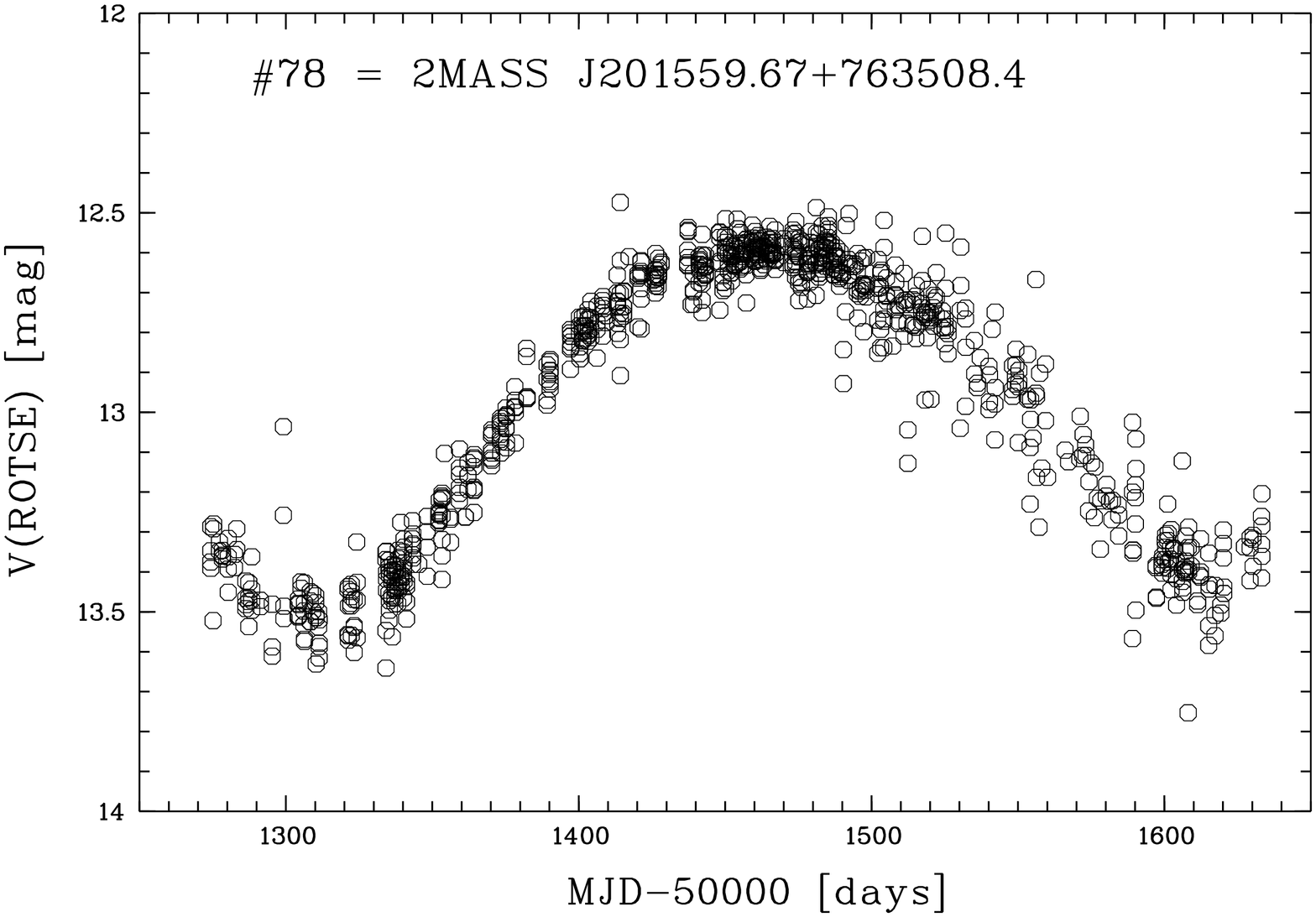}}}
\resizebox{8.5cm}{!}{\rotatebox{-90}{\includegraphics{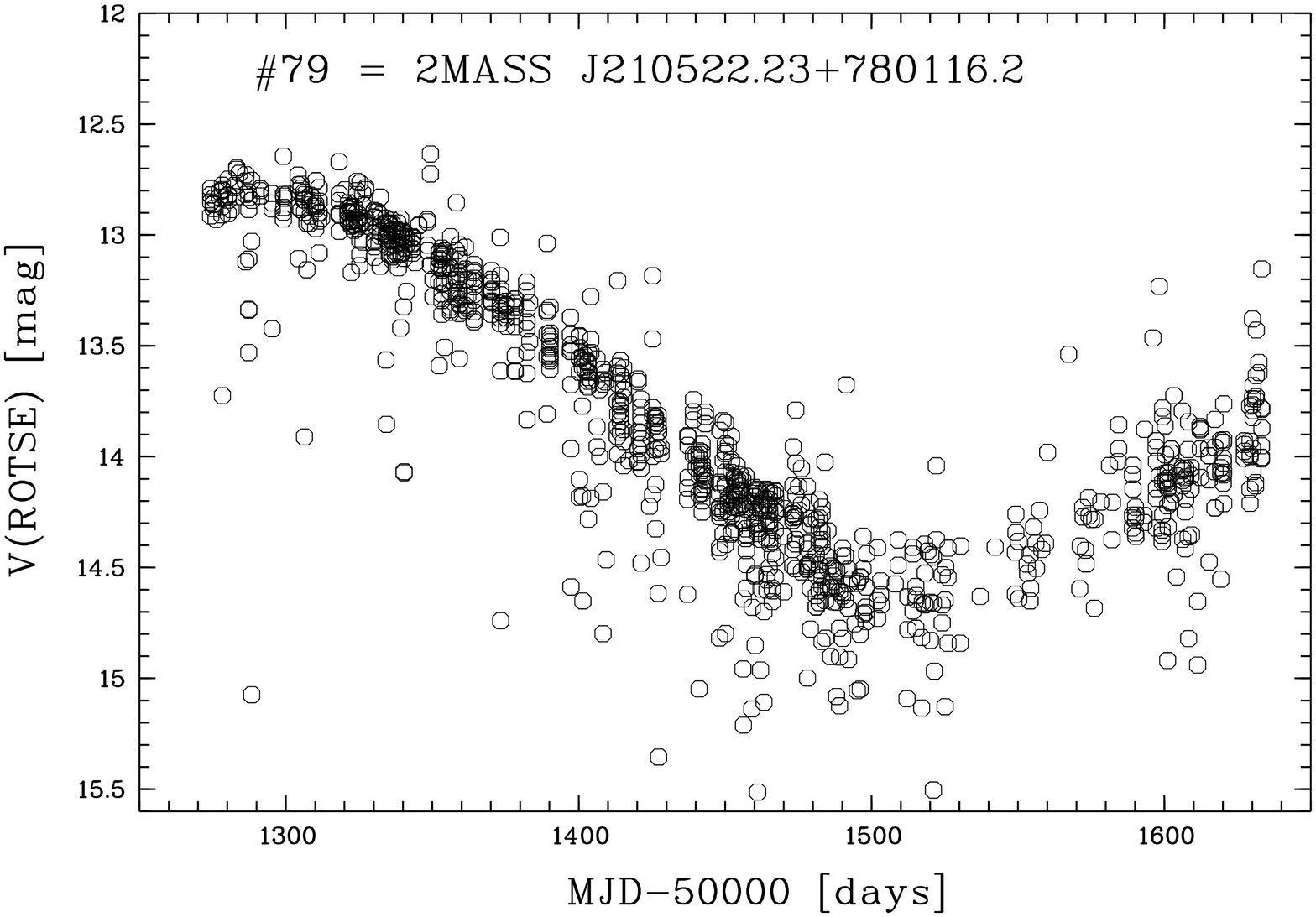}}}
\resizebox{8.5cm}{!}{\rotatebox{-90}{\includegraphics{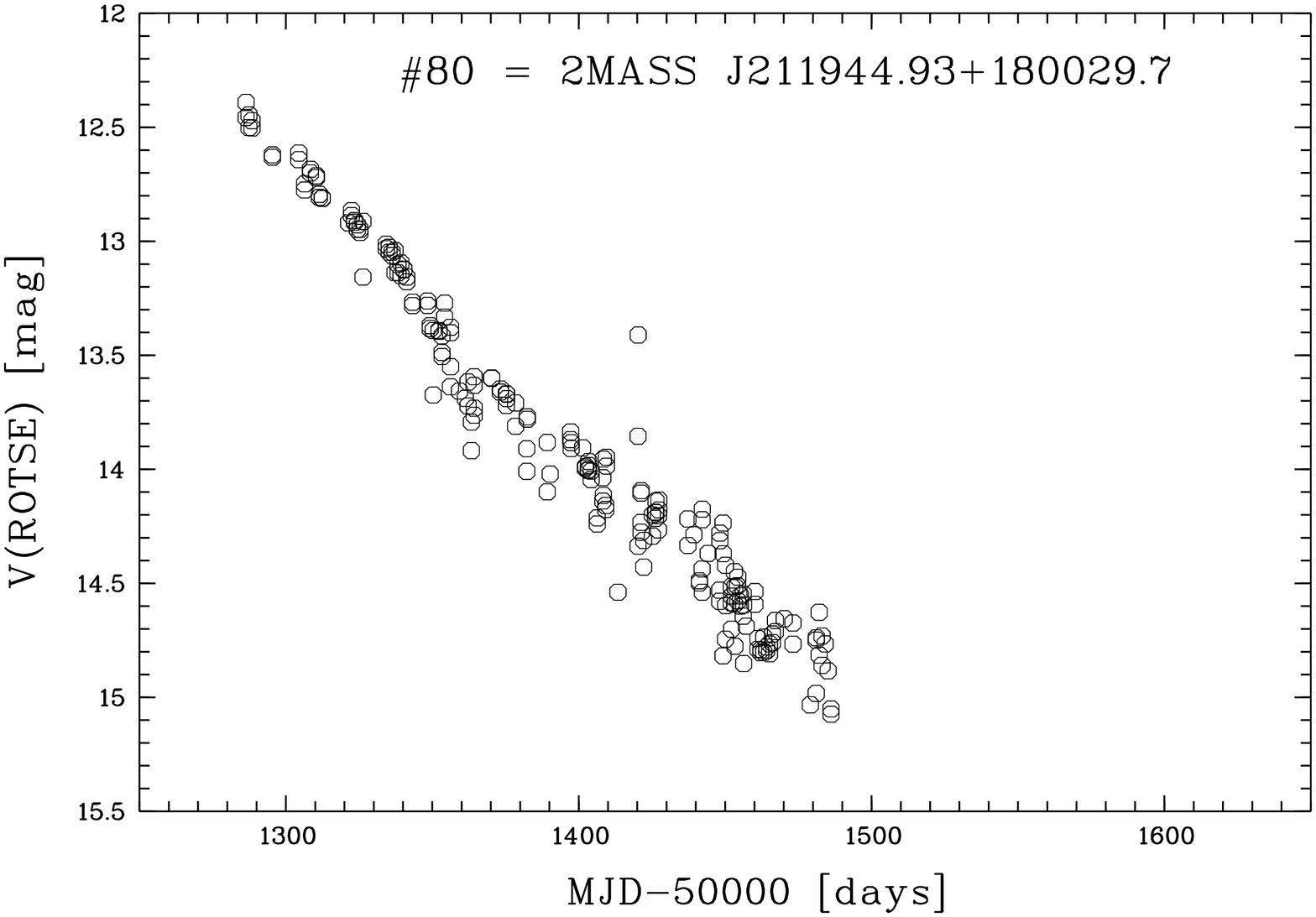}}}
\resizebox{8.5cm}{!}{\rotatebox{-90}{\includegraphics{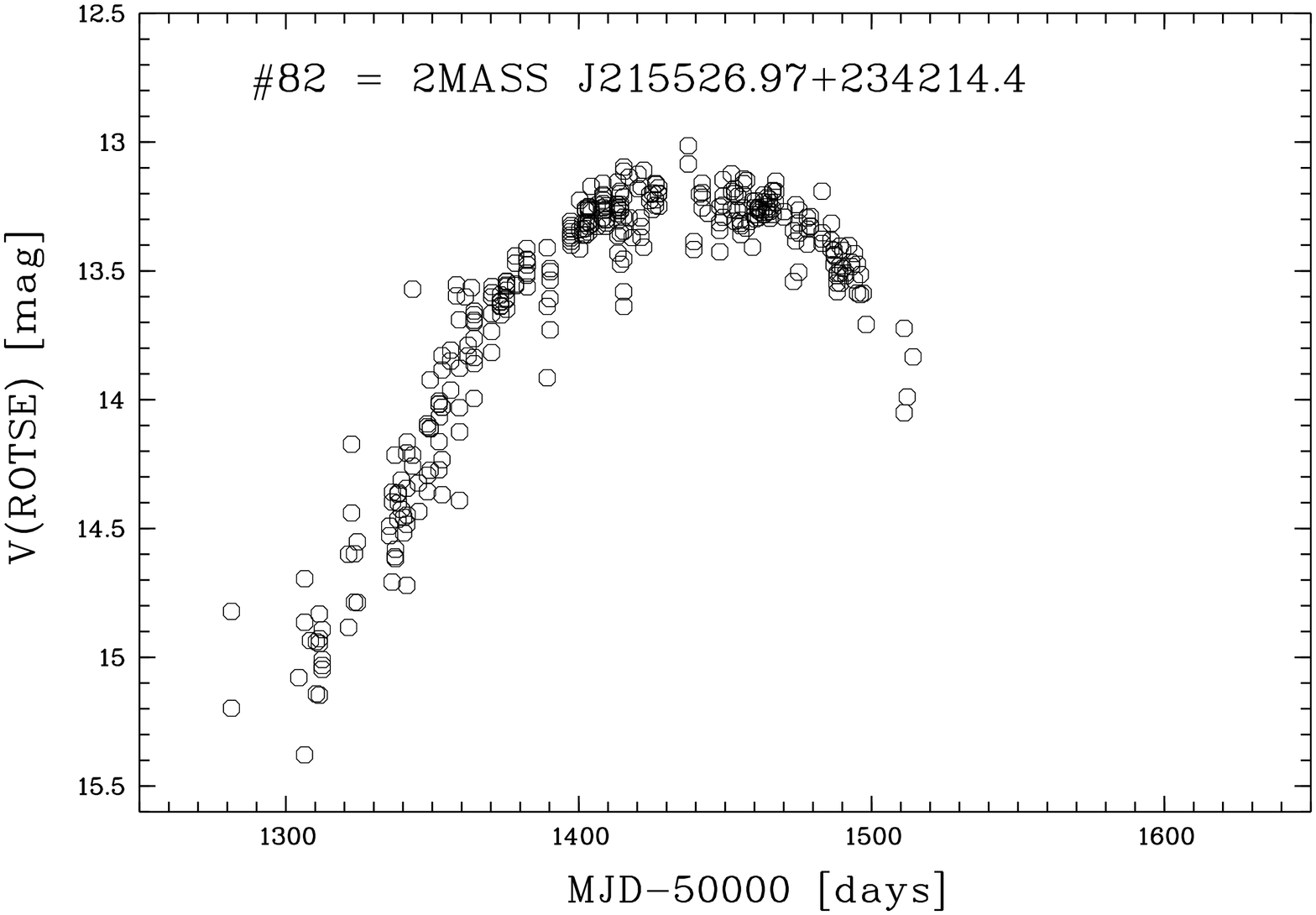}}}
\resizebox{8.5cm}{!}{\rotatebox{-90}{\includegraphics{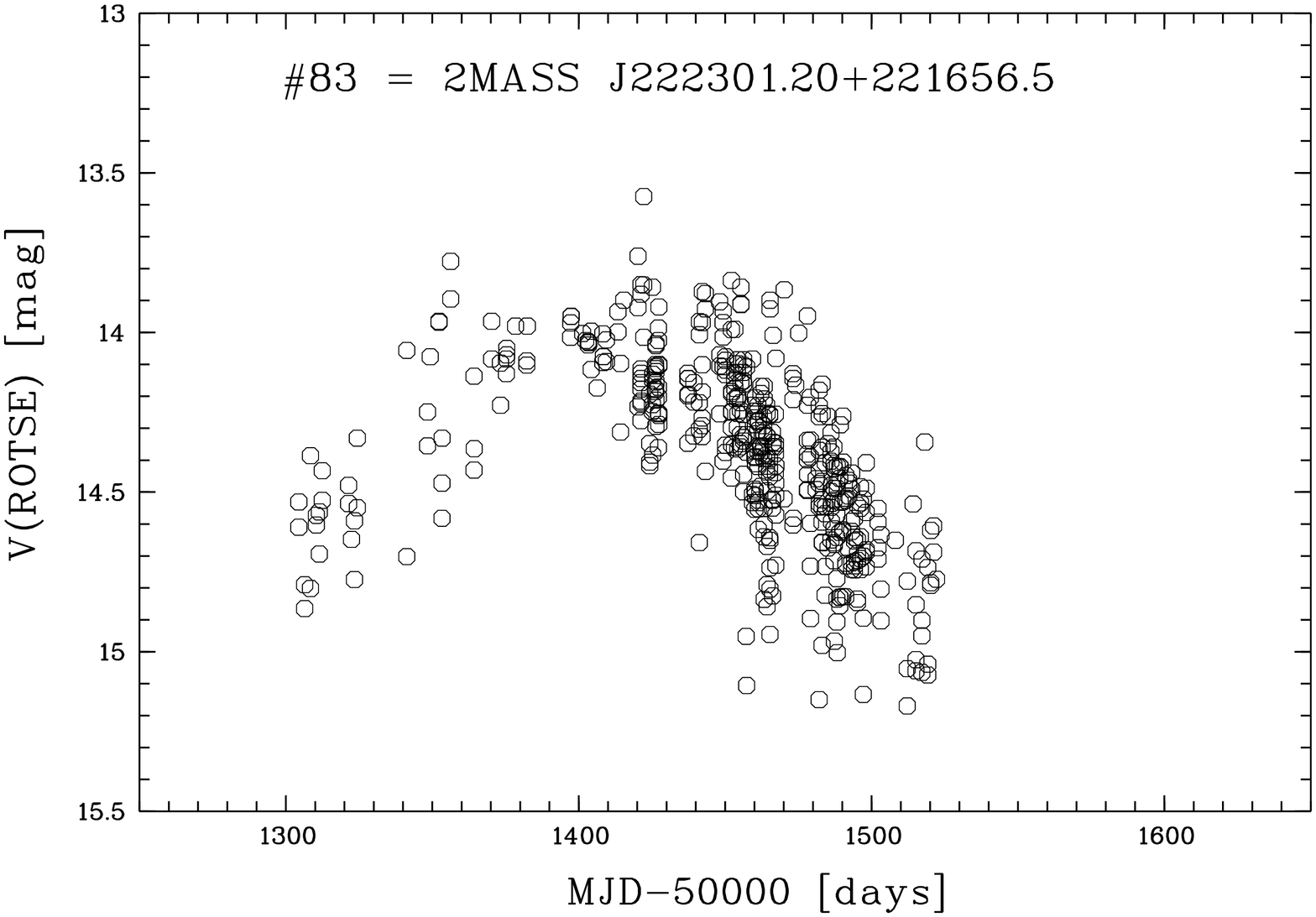}}}
\end{figure*}

\end{document}